\documentclass[%
 reprint,
nofootinbib,
 amsmath,amssymb,
 aps,
floatfix,
]{revtex4-1}

\usepackage{graphicx}
\usepackage{dcolumn}
\usepackage{bm}
\usepackage{amssymb}	
\usepackage{amsthm}	
\usepackage{feynmp}
\usepackage[sc]{mathpazo}
\usepackage{mathrsfs}
\usepackage{booktabs}
\usepackage{rotating}
\usepackage[flushleft]{threeparttable}
\usepackage{amsmath}
\makeatletter
\renewcommand*\env@matrix[1][*\c@MaxMatrixCols c]{%
  \hskip -\arraycolsep
  \let\@ifnextchar\new@ifnextchar
  \array{#1}}
\makeatother

\begin{document}

\preprint{APS/123-QED}

\title{Single pion production in neutrino-nucleon Interactions}

\author{M. Kabirnezhad}%
 \email{monireh.kabirnezhad@ncbj.gov.pl}
\affiliation{%
National Centre for Nuclear Research }%

\date{\today}

\begin{abstract}
This work represents an extension of the single pion production model proposed by Rein \cite{Rein}. The model consists of resonant
pion production and nonresonant background contributions coming from three Born diagrams in the helicity basis. The new work includes lepton
mass effects, and nonresonance interaction is described by five diagrams based on a nonlinear $\sigma$ model.
This work provides a full kinematic description of single pion production in the neutrino-nucleon interactions including resonant and nonresonant interactions, in the helicity basis, in order to study the interference effect.
\end{abstract}

\pacs{Valid PACS appear here}
\maketitle


\section{\label{intro}Introduction}
Neutrino-nucleon interactions that produce a single pion in the final state are of critical importance to accelerator-based neutrino experiments.
These single pion production (SPP) channels make up the largest fraction of the inclusive neutrino-nucleus cross section in the 1-3 GeV range, a region covered by most accelerator-based neutrino beams. The NuMI (NO$\nu$A) and proposed LBNF (DUNE) beams \cite{NuMI,DUNE} both peak near $2\text{ GeV}$, while the lower energy T2K and BNB \cite{T2K,BNB} beams have a significant portion of their flux in this region. \\
Models of SPP cross section processes are required to accurately predict the number and topology of observed charged-current (CC) neutrino interactions, and to estimate the dominant source of neutral-current (NC) backgrounds, where a charged (neutral) pion is confused for a final-state muon (electron). These experiments make use of nuclear targets. The foundation of neutrino-nucleus interaction models are neutrino-nucleon reaction processes like the one described in this paper.\\
Single pion production from a single nucleon occurs when the exchange boson has the requisite four-momentum to excite the target nucleon to a resonance state which promptly decays to produce a final-state pion (resonant interaction), or to create a pion at the interaction vertex (nonresonant interaction).
These interactions are distinguished from the lower four-momentum exchange quasielastic (QE) processes by the production of a final-state pion. However, they still resolve the nucleon as a whole, unlike the higher four-momentum exchange deep-inelastic scattering (DIS) interactions which interact with the nucleon's constituent quarks. \\
The SPP processes have been modeled in the $\Delta$ resonance region ($W<1.4~\text{GeV}$, where $W$ is invariant mass) \cite{Adler,HNV,Sato}, and updated to include more isospin $\frac{1}{2}$ resonance states \cite{Fogli,Alam}. However, models for neutrino interaction generators such as NEUT (the primary neutrino interaction generator used by the T2K experiment) \cite{NEUT} require that all resonances up to $W=2~\text{GeV}$ be included to accurately predict neutrino interaction rates. \\
The Rein and Sehgal (RS) model \cite{RS} does include these higher resonances,
but does not include a reliable model for nonresonant processes and related interference terms, and also neglects lepton mass effects.
NEUT and GENIE use the RS model for SPP by default, although they have made minor tweaks and improvements to their implementations,
like NEUT includes charged lepton masses \cite{BS} and a new form factor \cite{GS}.
In a later paper~\cite{Rein} Rein suggests how to coherently include the helicity amplitudes of the nonresonant contribution to the helicity amplitudes of the original RS model which is derived from a relativistic quark model \cite{FKR}. This update still neglects lepton mass effects.\\
In this work, we improve upon the ideas put forth by Rein by incorporating the nonresonant interactions introduced by Hernandez, Nieves, and Valverde (the HNV model) \cite{HNV}.
The previously neglected lepton mass effects, as well as several other features that make this model suitable for neutrino generators, are also included.\\
The resulting model has a full kinematic description of the final state particles, including pion angles, for CC
neutrino-nucleon and antineutrino-nucleon interactions,
\noindent\begin{minipage}{.4\linewidth}
\begin{eqnarray}\label{C2_29}
\nu_{\mu} + p \rightarrow & \mu p \pi^{+}~,\label{PP}\nonumber\\
\nu_{\mu} + n \rightarrow &\mu p \pi^{0}~, \label{NN}\nonumber\\
\nu_{\mu} + n \rightarrow &\mu n \pi^{+}~, \label{NP}\nonumber
\end{eqnarray}
\end{minipage}%
\begin{minipage}{.7\linewidth}
\begin{eqnarray}\label{CCanu_chan}
\bar{\nu}_{\mu} + n \rightarrow &\mu^{+} n \pi^{-}~,\nonumber\\
\bar{\nu}_{\mu} + p \rightarrow &\mu^{+} n \pi^{0}~, \nonumber\\
\bar{\nu}_{\mu} + P \rightarrow &\mu^{+} p \pi^{-}~, \nonumber
\end{eqnarray}
\end{minipage}
\\
\\
as well as for NC neutrino-nucleon and antineutrino-nucleon interactions:\\
\noindent\begin{minipage}{.4\linewidth}
\begin{eqnarray}
\nu + p \rightarrow &\nu p \pi^{0}~,\nonumber\\
\nu + p \rightarrow &\nu n \pi^{+}~, \nonumber\\
\nu + n \rightarrow &\nu n \pi^{0}~ ,\nonumber\\
\nu + n \rightarrow &\nu p \pi^{-}~ ,\nonumber
\end{eqnarray}
\end{minipage}%
\begin{minipage}{.7\linewidth}
\begin{eqnarray}\label{NCanu_chan}
\bar{\nu} + p \rightarrow &\bar{\nu} p \pi^{0}~,\nonumber\\
\bar{\nu} + p \rightarrow &\bar{\nu} n \pi^{+}~, \nonumber\\
\bar{\nu} + n \rightarrow &\bar{\nu} n \pi^{0}~,\nonumber\\
\bar{\nu} + n \rightarrow &\bar{\nu} p \pi^{-}. \nonumber
\end{eqnarray}
\end{minipage}
\section{General framework}\label{framework}
\begin{figure}[t]
\centering
\includegraphics [width=0.6\linewidth]{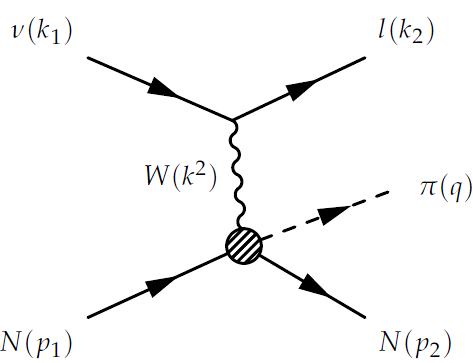}
\caption{Single pion production off nucleons}\label{SSP}
\end{figure}

\begin{figure}[t]
\centering
\includegraphics [width=0.8\linewidth]{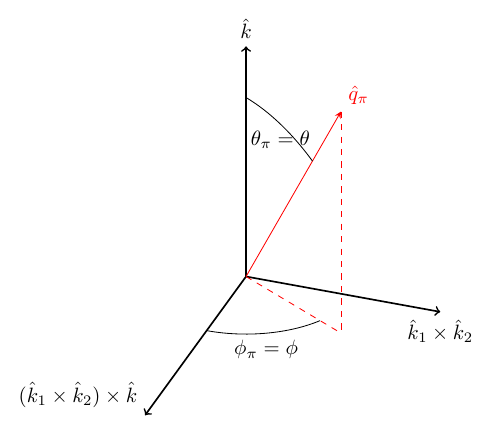}
\caption{Isobaric frame or the $\pi N $ center-of-mass frame.}\label{Isoframe}
\end{figure}

Single pion production in neutrino-nucleon interactions can be generally defined as:
\begin{equation}
\nu(k_1) + N(p_1) \longrightarrow l(k_2) N(p_2) \pi(q),\label{SPP_int}
\end{equation}
where $l$ is the outgoing charged lepton (neutrino) in CC (NC) interactions.
The diagram in Fig.~\ref{SSP} shows the momenta for each particle in the SPP interaction.
The incoming and outgoing lepton four-momenta are $k_1$  and $k_2$, respectively.
The nucleon four-momenta, similarly, are given by $p_1$ and $p_2$, and the final state pion four-momenta is denoted by $q$.
The momentum transfer is thus defined by $\mathbf{k} =  \mathbf{k}_1 - \mathbf{k}_2$, giving $Q^2= -k^2 = -(k_1 - k_2)^2$.\\
The transition amplitude for SPP (\ref{SPP_int}) can be written as
\begin{eqnarray}\label{trans_amp}
\mathcal{M}(\nu N \rightarrow l N' \pi) = \frac{G_F}{\sqrt{2}}~a~\epsilon^{\rho} ~\langle~N'\pi|~J_{\rho}~|N ~\rangle \nonumber\\
 \end{eqnarray}
 \unitlength=0.5mm
 where $\epsilon^{\rho}$ is leptonic current and $a$ is either the cosine of the Cabibbo angle for CC interactions or $1$ for NC interactions,
 \begin{eqnarray}
\epsilon^{\rho}_{CC}=& \bar{u}_{l}(k_2) \gamma^{\rho}(1- \gamma_5) u_{\nu}(k_1)\nonumber\\
\epsilon^{\rho}_{NC}=& \frac{1}{2} \bar{u}_{\nu}(k_2) \gamma^{\rho}(1- \gamma_5) u_{\nu}(k_1).
\label{epsilon}
\end{eqnarray}
While the hadronic currents for CC and NC interactions are different, they can both be decomposed into vector
and axial vector currents: $J_{\rho}= J^V_{\rho} - J^A_{\rho}$.\\
Calculations of the cross sections are simplified by working in the
isobaric (or Adler) frame. This is defined as the rest frame of the nucleon-pion system, where
 \begin{equation}
  \mathbf{q} + \mathbf{p}_2 = \mathbf{k} + \mathbf{p}_1=0 .
\end{equation}
As can be seen in Fig.~\ref{Isoframe},
when the momentum transfer is taken to be along the $\hat{z}$ axis in the Adler frame, the angle between the momentum transfer and pion direction can be used to define the polar ($\theta$) and azimuthal ($\phi$) angles of the pion.
\subsection{Lepton current}
For CC interactions the outgoing charged lepton is massive, while in the NC case it is massless and $ k_{02} = \mathbf{k}_2$.
The massive lepton of the CC case can have both right-handed and left-handed helicities, and the lepton current can be defined as:
\begin{equation}
\epsilon_{\lambda}^{\rho}= \bar{u}_{l _{\lambda}}(k_2) \gamma^{\rho}(1- \gamma_5) u_{\nu _L}(k_1), \label{epsilonLR}
\end{equation}
where $\lambda = - (+)$ for a left-handed (right-handed) lepton.
The components of the lepton current are thus related to $\lambda$, and when expressed in the isobaric frame as shown in Fig.~\ref{Isoframe}, they are
\begin{eqnarray}
\epsilon^0_{\lambda} &=& 2\lambda  A_{\lambda} \sqrt{1-\lambda cos\delta}~,\nonumber\\
\epsilon^1_{\lambda} &=& 2\lambda  A_{\lambda} \frac{k_{01} -\lambda |\bf{K_2}|}{|\bf{k}|} \sqrt{1+\lambda cos\delta}~,\nonumber\\
\epsilon^2_{\lambda} &=& 2i  A_{\lambda} \sqrt{1+\lambda cos\delta}~,\nonumber\\
\epsilon^3_{\lambda} &=& 2\lambda  A_{\lambda} \frac{k_{01} +\lambda |\bf{K_2}|}{|\bf{k}|} \sqrt{1-\lambda cos\delta}~,\label{epsilon_com}
\end{eqnarray}
where
\begin{equation}\label{}
  A_{\pm}= \sqrt{ k_{01}(k_{02} \mp |\mathbf{k}_2|)}.
\end{equation}
The neutrino energy is $k_{01}$, and the angle between the neutrino and the charged lepton in the $N\pi$ rest frame is denoted by $\delta$.
For simplicity, the $xz$ plane is defined such that lepton momentum $k_{1y} = k_{2y} =0$.\\
The lepton current $\epsilon^{\rho}$ can be interpreted as the intermediate gauge boson's polarization vector
\begin{eqnarray}
\epsilon^{\rho}_{\lambda} = \left[C_{L_{\lambda}} e^{\rho}_{L} + C_{R_{\lambda}} e^{\rho}_{R} + C_{\lambda} e^{\rho}_{\lambda} \right],
\end{eqnarray}
where $\mathbf{e}_L$ and $\mathbf{e}_R$ are the transverse polarizations (i.e., perpendicular to the momentum transfer),
and $\mathbf{e}_{\lambda}$ is the longitudinal polarization which is along the $\mathbf{z}$ direction of the isobaric system. This gives
\begin{eqnarray}\label{pol_vec}
e^{\alpha}_{L} &=& \frac{1}{\sqrt{2}} \begin{pmatrix} 0&1 &-i&0\end{pmatrix} ~,\nonumber\\
e^{\alpha}_{R} &=& \frac{1}{\sqrt{2}} \begin{pmatrix} 0&-1&-i&0 \end{pmatrix}~, \nonumber\\
e^{\alpha}_{\lambda} &=& \frac{1}{\sqrt{|(\epsilon^0_{\lambda})^2 - (\epsilon^3_{\lambda})^2 |}}  \begin{pmatrix} \epsilon^0_{\lambda}&0&0&\epsilon^3_{\lambda}\end{pmatrix} \label{W_pol}\nonumber\\
\end{eqnarray}
and,
\begin{eqnarray}
C_{L_{\lambda}} &=&  \frac{1}{\sqrt{2}} \left(\epsilon_{\lambda}^{1}+ i\epsilon_{\lambda}^{2} \right)~, \nonumber\\
C_{R_{\lambda}} &=&  - \frac{1}{\sqrt{2}} \left(\epsilon_{\lambda}^{1}- i\epsilon_{\lambda}^{2} \right)~,\nonumber\\
C_{\lambda} &=& \sqrt{|(\epsilon^0_{\lambda})^2 - (\epsilon^3_{\lambda})^2 |}~.
\end{eqnarray}
\subsection{Hadron currents}
 Hadronic currents can be decomposed into vector and axial vector parts:
\begin{equation}
\langle~N\pi|~J^{\rho}~|N ~\rangle = \langle~N\pi|~J_V^{\rho}- J_A^{\rho}~|N ~\rangle~.
\end{equation}
We can further decompose the vector and axial vector parts as
\begin{eqnarray}\label{4hadron}
{J^{\rho}_{V}} e_{\rho}^{\lambda_k}  &= \sum_{k=1}^{6} V_k(s,t,u)~\bar{u}_N(p_{2})O^{\lambda_k}(V_k) u_N(p_{1})~,\label{Dirac_current}\nonumber\\
{J^{\rho}_{A}} e_{\rho}^{\lambda_k} &= \sum_{k=1}^{8} A_k(s,t,u)~\bar{u}_N(p_{2})O^{\lambda_k}(A_k) u_N(p_{1})~,\nonumber\\
\end{eqnarray}
where $\lambda_k$ stands for the gauge boson's polarization, $e_{L},~e_R$ or $e_{\pm}$.
The Dirac equation allows for $16$ independent Lorentz invariants $O(V_k)$ and $O(A_k)$. However, vector current conservation reduces the number of $O(V_k)$ to six. Lorentz invariants are given in Ref.~\cite{Adler} and can also be found in Appendix \ref{appA}.
Invariant amplitudes $V_k$ and $A_k$ can be calculated once the interactions and their associated diagrams are defined. They are generally a function of the following invariant Mandelstam variables:
\begin{eqnarray}
s = (p_2 + q)^2 = (p_1 + k)^2 = W^2~,\nonumber\\
t = (k-q)^2~, ~~ \text{and}~~ u=(q-p_1)^2~.
\end{eqnarray}
Using the representations of Dirac matrices and spinors in terms of two-dimensional Pauli matrices and spinors,
 we can rewrite the right-hand side of Eq.~(\ref{Dirac_current}) in terms of $2\times2$ matrices $\Sigma_k$ and $\Lambda_k$,
\begin{eqnarray}\label{isobar_dicom}
 {J^{\rho}_{V}} e_{\rho}^{\lambda_k} &= \sum_{k=1}^{6} \mathscr{F}_k(s,t,u)~\chi_2^{\ast}~\Sigma^{\lambda_k}_k ~ \chi_1~,\nonumber\\
{J^{\rho}_{A}} e_{\rho}^{\lambda_k}  &= \sum_{k=1}^{8} \mathscr{G}_k(s,t,u)~\chi_2^{\ast}~\Lambda^{\lambda_k}_k ~ \chi_1~,
\end{eqnarray}
where $\chi_1$ ($\chi_2$) is the Pauli spinor of the incident (outgoing) nucleon.\\
Definitions for $\Sigma_k$ and $\Lambda_k$ as well as $\mathscr{F}_k$ and $\mathscr{G}_k$ which are related to the invariant amplitudes, are given in Appendix \ref{appA}.
\subsection{Helicity amplitudes}
  Helicity amplitude can be defined with three indices: incident nucleon helicity ($\lambda_1$), outgoing nucleon helicity ($\lambda_2$), and gauge boson's polarization (pions are spinless).
 From Eqs.~(\ref{epsilonLR}) and (\ref{trans_amp}), we have
\begin{align}\label{M_HA1}
&\mathcal{M}_{CC}(\nu N \rightarrow l_{\lambda} N' \pi)
= \frac{G_F}{\sqrt{2}}\cos \theta_C ~\langle~N'\pi| ~\epsilon_R^{\rho}J_{\rho} ~|N ~\rangle \nonumber\\
=& \frac{G_F}{\sqrt{2}}\cos \theta_C ~\langle~N'\pi| ~C_{L_{\lambda}} e^{\rho}_{L}J_{\rho} + C_{R_{\lambda}}e^{\rho}_{R}J_{\rho} + C_{\lambda}e^{\rho}_{\lambda}J_{\rho}|N ~\rangle~,
 \end{align}
  where there are four independent gauge boson's polarizations from Eq.~(\ref{pol_vec}), i.e.,$e_{L},~e_R$ and $e_{\pm}$.
 Using Eq.~(\ref{M_HA1}), we can define the helicity amplitudes for vector and axial currents:
 \begin{align}\label{HA_definition}
\tilde{F}_{\lambda_2, \lambda_1}^{\lambda_k} &= \langle~N\pi|~e^{\rho}_{\lambda_k} V_{\rho}~|N ~\rangle~,\nonumber\\
\tilde{G}_{\lambda_2, \lambda_1}^{\lambda_k} &= \langle~N\pi|~e^{\rho}_{\lambda_k} A_{\rho}~|N ~\rangle~,
\end{align}
where $\lambda_k$ stands for gauge boson's polarizations and
\begin{align}\label{}
V= \frac{1}{2M} J^V ~~,~~~~~~~~~~~~ A= \frac{1}{2M} J^A~.
\end{align}
For each vector and axial current, we can define $2\times2\times4=16$ helicity amplitudes, $\tilde{F}^{(\lambda_k)}_{\lambda_2,\lambda_1}$ and $\tilde{G}^{(\lambda_k)}_{\lambda_2,\lambda_1}$, respectively.
The final results for all helicity amplitudes are summarized in Table.~\ref{helicity_amp}.
\subsection{Cross section}
A general form of the differential cross section for single pion production is\footnote{$\tilde{F}$ ($\tilde{G}$) is a function of $E,~W,~Q^2,~\theta$, and $\phi$, but here we only show $\theta$ and $\phi$ in comparison with $F^j$ ($G^j$) in Eq.~(\ref{Mulex}), which is not a function of pion angles.}
\begin{widetext}
\begin{align}
&\frac{d\sigma(\nu N  \rightarrow lN\pi)}{dk^2 dW d\Omega_{\pi}} =
\frac{G_F^2}{2} \frac{1}{(2\pi)^4} \frac{|\bf{q}|}{4} \frac{-k^2}{(k^L)^2}\sum_{\lambda_2, \lambda_1}\bigg\{ \nonumber\\
\Big|& C_{L_-} (\tilde{F}_{\lambda_2 \lambda_1}^{e_L}(\theta,\phi)- \tilde{G}_{\lambda_2 \lambda_1}^{e_L}(\theta,\phi))
+ C_{R_-} (\tilde{F}_{\lambda_2 \lambda_1}^{e_R}(\theta,\phi) - \tilde{G}_{\lambda_2 \lambda_1}^{e_R}(\theta,\phi))
+ C_{-} (\tilde{F}_{\lambda_2 \lambda_1}^{e_{-}}(\theta,\phi) - \tilde{G}_{\lambda_2 \lambda_1}^{e_{-}}(\theta,\phi))\Big|^{2}\nonumber\\
+ \Big|& C_{L_+} (\tilde{F}_{\lambda_2 \lambda_1}^{e_L}(\theta,\phi) - \tilde{G}_{\lambda_2 \lambda_1}^{e_L}(\theta,\phi))
+ C_{R_+} (\tilde{F}_{\lambda_2 \lambda_1}^{e_R}(\theta,\phi)- \tilde{G}_{\lambda_2 \lambda_1}^{e_R}(\theta,\phi))
+ C_{+} (\tilde{F}_{\lambda_2 \lambda_1}^{e_+}(\theta,\phi)- \tilde{G}_{\lambda_2 \lambda_1}^{e_+}(\theta,\phi))\Big|^{2} \bigg\}.
\label{Xsec}
\end{align}
\end{widetext}
For antineutrino interactions, one needs to swap $C_{L_{\pm}}$ with $C_{R_{\pm}}$.
 An equivalent differential cross section with an explicit form for the angle $\phi$ is given in Appendix \ref{xsec_app}.
\subsubsection{Multipole expansion} \label{MultipoleExpansion}
Helicity amplitudes are invariant under ordinary rotation; therefore, 
it is always possible to expand them over angular momenta \cite{wick,jackson}.
To do this first we need to have a standard\footnote{The helicity amplitudes in Eq.~(\ref{HA_definition}) are not independent.} form for helicity amplitudes \cite{Rein}:
\begin{eqnarray} \label{standard_HA}
F_{\mu\lambda}(\theta,\phi), ~~~~ G_{\mu\lambda}(\theta,\phi)
\end{eqnarray}
with two indexes
\begin{eqnarray}
\lambda&= \lambda_k - \lambda_1 , ~~~~~~~~ \lambda= -\frac{3}{2}, -\frac{1}{2}, \frac{1}{2}, \frac{3}{2}\nonumber\\
\mu&= \lambda_q - \lambda_2 = -\lambda_2, ~~~~~~ \mu=-\frac{1}{2}, \frac{1}{2} ,
     \end{eqnarray}
where $\lambda_k$ is the polarization of the gauge bosons; $\lambda_k(e_L)=-1$, $\lambda_k(e_R)=+1$, and $\lambda_k(e_{\pm})=0$. The helicity of the pion, $\lambda_q$, is zero.\\
There is a simple relation between the standard helicity amplitudes of Eq.~(\ref{standard_HA}) and the helicity amplitudes used in Eq.~(\ref{Xsec}):
\begin{align}\label{stantoHA1}
F_{\mu\lambda}(\theta,\phi) &= e^{i[\lambda_1\pi + \lambda_2(\pi + 2\phi)]} \tilde{F}^{\lambda_k}_{\lambda_2,\lambda_1}(\theta,\phi),\nonumber\\
G_{\mu\lambda}(\theta,\phi) &= e^{i[\lambda_1\pi + \lambda_2(\pi + 2\phi)]} \tilde{G}^{\lambda_k}_{\lambda_2,\lambda_1}(\theta,\phi).
\end{align}
The standard helicity amplitudes allow for the use of multipole expansion \cite{wick, Rein}:
 \begin{align}\label{Mulex}
F_{\mu\lambda}(\theta,\phi)&=&\sum_j F^j_{\mu\lambda}(2j+1) d^j_{\lambda\mu}(\theta) e^{i(\lambda - \mu)\phi}\nonumber\\
G_{\mu\lambda}(\theta,\phi)&=&\sum_j G^j_{\mu\lambda}(2j+1) d^j_{\lambda\mu}(\theta) e^{i(\lambda - \mu)\phi},
\end{align}
where $ \sqrt{\frac{2j+1}{4\pi}}d^j_{\lambda,\mu}(\theta)e^{i(\lambda - \mu)\phi}$ are mutually orthonormal functions \cite{wick}.
The same multipole expansion can be used for $F^{(\pm)0}_{\mu\lambda}$ ($G^{(\pm)0}_{\mu\lambda}$).
\section{Resonance Contribution and Nonresonant Background} \label{Res_bg}
\subsection{Single pion production via resonance decay}
The RS-model \cite{RS} describes SPP in neutrino-nucleon interaction via resonance decay, and it is based on helicity amplitudes derived from a relativistic
quark model \cite{FKR}. The quark model had been extended to neutrino interactions by Ravndal \cite{Ravndal}.
The original RS-model \cite{RS} includes 18 resonances up to $M_R \leq 2~\text{GeV}$. However, according to Ref.~\cite{PDG} one is no longer in use. The remaining $17$ are given in Table~\ref{res_table}.
The RS-model also neglected the mass of the charged lepton but it has been restored in Refs. \cite{russa,BS,kris}.
\begin{table}[h]
\centering
\caption{Nucleon-resonances below $2 GeV/c^2$  }
\label{res_table}
\renewcommand{\arraystretch}{1.3}
\begin{ruledtabular}
 \begin{tabular}{lccccl}
 Resonance & $M_R$ &$\Gamma_0$& $\chi_E$&$ \sigma^D$ &$N$   
 \\ [0.1ex]
 \hline
 $P_{33}(1232)$ & 1232 & 117& 1   & + &0 \\
$ P_{11}(1440)$ & 1430 & 350& 0.65& +  &2\\
 $D_{13}(1520)$ &1515  & 115& 0.60& -  &1\\
 $S_{11}(1535)$ & 1535&150 &  0.45& -  &1\\
  $P_{33}(1600)$ & 1600 & 320& 0.18& + &2\\
 $S_{31}(1620)$ & 1630 &140& 0.25 & + &1\\
  $S_{11}(1650)$ &1655&140& 0.70  & + &1\\
  $D_{15}(1675)$ &1675&150& 0.40  & +  &1\\
  $F_{15}(1680)$&1685&130 & 0.67  & +  &2\\
  $D_{13}(1700)$&1700&150& 0.12   & -  &1\\
  $D_{33}(1700)$&1700&300& 0.15   & +  &1\\
$  P_{11}(1710)$&1710&100& 0.12   & -  &2\\
$  P_{13}(1720)$ & 1720&250& 0.11 & +  &2\\
$ F_{35} (1905)$ & 1880&330& 0.12 & -  &2\\
$ P_{31} (1910)$ & 1890&280 &0.22 & -  &2\\
$ P_{33}(1920)$ &  1920&260 & 0.12& +  &2\\
$ F_{37}(1950)$ & 1930&285 &0.40  & +  &2\\
\end{tabular}
\end{ruledtabular}
\begin{tablenotes}
  \item Resonances are identified with isospin ($I$) and angular momentum ($j$); $L_{2I,2j}$. The Breit-Wigner (BW) mass ($M_R$[MeV]), BW full width ($\Gamma_0$[MeV]) and $\pi N$ branching ratio ($\chi_E$) are reproduced from Ref.~\cite{PDG}. The decay signs ($\sigma^D$) and the number of oscillators are from Ref.~\cite{RS}.
    \end{tablenotes}
\end{table}
The helicity amplitudes in Ref.~\cite {RS} are referring to a single resonance with a well-defined angular momentum, isospin and helicity. Each helicity amplitude defines a specific resonance and its subsequent decay into the $N\pi$ final state:
\begin{align}
&\langle N\pi,\lambda_2| \epsilon^{\alpha} J_{\alpha}| N,\lambda_1\rangle\nonumber\\
=& \langle N\pi,\lambda_2|R \lambda_R \rangle \langle R\lambda_R|\epsilon^{\alpha} J_{\alpha}|N \lambda_1 \rangle.
\end{align}
For the vector component of resonant production, we have
\begin{align}
&\tilde{F}_{\lambda_1, \lambda_2} ^{\lambda_k}(\theta, \phi) = e^{-i[\lambda_1\pi + \lambda_2(\pi + 2\phi)]}F_{\mu,\lambda}(\theta,\phi)\nonumber\\
=& e^{-i[\lambda_1\pi + \lambda_2(\pi + 2\phi)]}\nonumber\\
 &\sum_j F^j_{\mu,\lambda} (2j+1) d^j_{\lambda,\mu}(\theta,\phi)e^{i(\lambda - \mu)\phi}~,
\end{align}
and similarly for the axial component ($\tilde{G}_{\lambda_1, \lambda_2} ^{\lambda_k}(\theta, \phi)$). The forms of
$F^j_{\mu,\lambda}$ and $G^j_{\mu,\lambda}$ are given in \cite{Rein}.\\
According to the RS model \cite{RS}, the decay amplitudes are
\begin{align}\label{D_amp}
\mathcal{D}^j(R)=\langle N\pi,\lambda_2|R \lambda_R \rangle &= \sigma^D C_{N\pi}^{j} \sqrt{\chi_E} \kappa C_{N\pi}^{I} f_{BW} 
\end{align}
where $\sigma^D$ and $\chi_E$ are given in Table~\ref{res_table} and $C^I_{N\pi}$ are the isospin Clebsch-Gordan coefficients given in Table~\ref{RSisospin} for CC and NC interactions.
\begin{table}
\centering
\caption{Isospin coefficients for RS model CC and NC (anti)neutrino channels. } \label{RSisospin}
\renewcommand{\arraystretch}{1.5}
\begin{ruledtabular}
 \begin{tabular}{|l|l|c|c|}
 {$\nu$ Channels}  & {$\bar{\nu}$ Channels} & $C^{3/2}_{N\pi}$& $C^{1/2}_{N\pi}$   \\ [0.7ex]
 \hline
$\nu p \rightarrow l^- p \pi^{+}$  &  $\bar{\nu} n \rightarrow l^+ n \pi^-$   & $\sqrt{3}$ & $0$\\
$\nu n \rightarrow l^- p \pi^{0}$  &  $\bar{\nu} p \rightarrow l^+ n \pi^0$   &  $-\sqrt{\frac{2}{3}}$  &  $\sqrt{\frac{1}{3}}$\\
$\nu n \rightarrow l^- n \pi^{+}$  &  $\bar{\nu} p \rightarrow l^+ p \pi^-$   &$\sqrt{\frac{1}{3}}$&$\sqrt{\frac{2}{3}}$\\
\hline
$\nu p \rightarrow \nu p \pi^{0}$  &  $\bar{\nu} p \rightarrow \bar{\nu} p \pi^0$   &  $\sqrt{\frac{2}{3}}$  &  $-\sqrt{\frac{1}{3}}$\\
$\nu p \rightarrow \nu n \pi^{+}$  &  $\bar{\nu} p \rightarrow \bar{\nu} n \pi^+$   &$-\sqrt{\frac{1}{3}}$  &$-\sqrt{\frac{2}{3}}$\\
$\nu n \rightarrow \nu n \pi^{0}$  &  $\bar{\nu} n \rightarrow \bar{\nu} n \pi^0$   &  $\sqrt{\frac{2}{3}}$  &  $\sqrt{\frac{1}{3}}$\\
$\nu n \rightarrow \nu p \pi^{-}$  &  $\bar{\nu} n \rightarrow \bar{\nu} p \pi^-$   &$\sqrt{\frac{1}{3}}$  &$-\sqrt{\frac{2}{3}}$\\
\end{tabular}
\end{ruledtabular}
\end{table}
The signs of the angular momentum Clebsch-Gordan coefficients are denoted as $C_{N\pi}^{j}$ as they are defined in Ref. \cite{Rein}.\\
As it is explained in \cite{Rein}, the cross section given in \cite{RS} is slightly different from what was given in Eq.~(\ref{Xsec}). Therefore a factor, $\kappa$, is defined for the identification:
\begin{equation}
\kappa= \left( 2\pi^2 \frac{W^2}{M^2} ~.~\frac{2}{2j+1} ~ \frac{1}{|\mathbf{q}|} \right)^{\frac{1}{2}},
\end{equation}
and
\begin{eqnarray}
f_{BW}(R) = \sqrt{\frac{\Gamma_R}{2\pi} }\left( \frac{1}{W- M_R + i\Gamma_R/2} \right )
\end{eqnarray}
is the Breit-Wigner amplitude with
\begin{eqnarray}
\Gamma_R = \Gamma_0 (|\mathbf{q}(W)|/|\mathbf{q}(M_R)|)^{2l+1},
\end{eqnarray}
where $\Gamma_0$ and $M_R$ are given in Table~\ref{res_table}. \\
The helicity amplitudes of the resonant interaction as a function of $\theta$ and $\phi$ are summarized in Table~\ref{res_HA} where $\mathcal{D}^j(R)$ is the decay amplitude given in Eq.~(\ref{D_amp}) and $f_{\substack{\pm1\\\pm3}}(R)$ and $f_{0\pm}^{(\pm)}(R)$ are given in \cite{RS} for both CC and NC neutrino interactions. \\
\begin{table*}
\centering
\caption{Helicity amplitudes of resonant interaction. }
\label{res_HA}
\renewcommand{\arraystretch}{1.3}
\begin{ruledtabular}
 \begin{tabular}{c|c|c|c}
  $\lambda_2$& $\lambda_1$ &$\tilde{F}_{\lambda_2 \lambda_1}^{e_{L}}(\theta, \phi)- \tilde{G}_{\lambda_2 \lambda_1}^{e_{L}}(\theta, \phi)$& $\tilde{F}_{\lambda_2 \lambda_1}^{e_{R}}(\theta, \phi)- \tilde{G}_{\lambda_2 \lambda_1}^{e_{R}}(\theta, \phi)$\\[4pt]
 \hline
 $\begin{aligned}
&\scalebox{0.001}{~}\\&\frac{1}{2}\\[7pt] -&\frac{1}{2} \\[7pt] &\frac{1}{2}\\[7pt] -&\frac{1}{2}
\end{aligned}$
&
 $\begin{aligned}
&\scalebox{0.001}{~}\\&\frac{1}{2}\\[7pt] &\frac{1}{2} \\[7pt] -&\frac{1}{2}\\[7pt] -&\frac{1}{2}
\end{aligned}$
&
 $\begin{aligned}
&\scalebox{0.001}{~}\\
-&\sum_j\frac{2j+1}{\sqrt{2}} \mathcal{D}^j(R) f_{+3}(R(I,j=l\pm\frac{1}{2}))  d^j_{\frac{3}{2} \frac{1}{2}}(\theta) e^{-2i\phi}\\
\mp&\sum_j\frac{2j+1}{\sqrt{2}} \mathcal{D}^j(R)~ f_{+3}(R(I,j=l\pm\frac{1}{2}))~  d^j_{\frac{3}{2} -\frac{1}{2}}(\theta)e^{-i\phi}\\
&\sum_j\frac{2j+1}{\sqrt{2}} \mathcal{D}^j(R) ~f_{+1}(R(I,j=l\pm\frac{1}{2}))~  d^j_{\frac{1}{2} \frac{1}{2}}(\theta)e^{-i\phi}\\
\pm&\sum_j\frac{2j+1}{\sqrt{2}} \mathcal{D}^j(R) ~f_{+1}(R(I,j=l\pm\frac{1}{2}))~  d^j_{\frac{1}{2} -\frac{1}{2}}(\theta)
\end{aligned}$
&
 $\begin{aligned}
 &\scalebox{0.01}{~}\\
-&\sum_j\frac{2j+1}{\sqrt{2}} \mathcal{D}^j(R)~ f_{-1}(R(I,j=l\pm\frac{1}{2}))~  d^j_{\frac{1}{2} -\frac{1}{2}}(\theta)\\
\pm&\sum_j\frac{2j+1}{\sqrt{2}} \mathcal{D}^j(R)~ f_{-1}(R(I,j=l\pm\frac{1}{2}))~  d^j_{\frac{1}{2} \frac{1}{2}}(\theta)e^{i\phi}\\
-&\sum_j\frac{2j+1}{\sqrt{2}} \mathcal{D}^j(R)~ f_{-3}(R(I,j=l\pm\frac{1}{2}))~  d^j_{\frac{3}{2} -\frac{1}{2}}(\theta)e^{i\phi}\\
\pm&\sum_j\frac{2j+1}{\sqrt{2}} \mathcal{D}^j(R) ~f_{-3}(R(I,j=l\pm\frac{1}{2}))~  d^j_{\frac{3}{2} \frac{1}{2}}(\theta) e^{2i\phi}
\end{aligned}$
\\\hline
&  &$\tilde{F}_{\lambda_2 \lambda_1}^{e_{-}}(\theta, \phi)- \tilde{G}_{\lambda_2 \lambda_1}^{e_{-}}(\theta, \phi)$& $\tilde{F}_{\lambda_2 \lambda_1}^{e_{+}}(\theta, \phi)- \tilde{G}_{\lambda_2 \lambda_1}^{e_{+}}(\theta, \phi)$\\[4pt]
 \hline
 $\begin{aligned}
&\scalebox{0.001}{~}\\&\frac{1}{2}\\[7pt] -&\frac{1}{2} \\[7pt] &\frac{1}{2}\\[7pt] -&\frac{1}{2}
\end{aligned}$
&
 $\begin{aligned}
&\scalebox{0.001}{~}\\&\frac{1}{2}\\[7pt] &\frac{1}{2} \\[7pt] -&\frac{1}{2}\\[7pt] -&\frac{1}{2}
\end{aligned}$
&
 $\begin{aligned}
&\scalebox{0.05}{~}\\
\mp&\frac{|\mathbf k|}{\sqrt{-k^2}}\sum_j\frac{2j+1}{\sqrt{2}} \mathcal{D}^j(R)~ f^{(-)}_{0-}(R(I,j=l\pm\frac{1}{2}))~  d^j_{-\frac{1}{2} -\frac{1}{2}}(\theta)e^{-i\phi}\\
-&\frac{|\mathbf k|}{\sqrt{-k^2}}\sum_j\frac{2j+1}{\sqrt{2}} \mathcal{D}^j(R)~ f^{(-)}_{0-}(R(I,j=l\pm\frac{1}{2}))~  d^j_{-\frac{1}{2} \frac{1}{2}}(\theta)\\
\pm&\frac{|\mathbf k|}{\sqrt{-k^2}}\sum_j\frac{2j+1}{\sqrt{2}} \mathcal{D}^j(R)~ f^{(-)}_{0+}(R(I,j=l\pm\frac{1}{2}))~  d^j_{\frac{1}{2} -\frac{1}{2}}(\theta)\\
-&\frac{|\mathbf k|}{\sqrt{-k^2}}\sum_j\frac{2j+1}{\sqrt{2}} \mathcal{D}^j(R)~ f^{(-)}_{0+}(R(I,j=l\pm\frac{1}{2}))~  d^j_{\frac{1}{2} \frac{1}{2}}(\theta)e^{i\phi}\end{aligned}$
&
 $\begin{aligned}
 &\scalebox{0.05}{~}\\
\mp&\frac{|\mathbf k|}{\sqrt{-k^2}}\sum_j\frac{2j+1}{\sqrt{2}} \mathcal{D}^j(R)~ f^{(+)}_{0-}(R(I,j=l\pm\frac{1}{2}))~  d^j_{-\frac{1}{2} -\frac{1}{2}}(\theta)e^{-i\phi}\\
-&\frac{|\mathbf k|}{\sqrt{-k^2}}\sum_j\frac{2j+1}{\sqrt{2}}\mathcal{D}^j(R)~ f^{(+)}_{0-}(R(I,j=l\pm\frac{1}{2}))~  d^j_{-\frac{1}{2} \frac{1}{2}}(\theta)\\
\pm&\frac{|\mathbf k|}{\sqrt{-k^2}}\sum_j\frac{2j+1}{\sqrt{2}} \mathcal{D}^j(R)~ f^{(+)}_{0+}(R(I,j=l\pm\frac{1}{2}))~  d^j_{\frac{1}{2} -\frac{1}{2}}(\theta)\\
-&\frac{|\mathbf k|}{\sqrt{-k^2}}\sum_j\frac{2j+1}{\sqrt{2}} \mathcal{D}^j(R)~ f^{(+)}_{0+}(R(I,j=l\pm\frac{1}{2}))~  d^j_{\frac{1}{2} \frac{1}{2}}(\theta)e^{i\phi}\end{aligned}$
\\
\end{tabular}
\end{ruledtabular}
\end{table*}
The resonance production amplitudes depend on the vector and axial form factors which have a dipole form in the RS-model. In this work we use the form factors proposed by Graczyk and Sobczyk (GS) in Reference \cite{GS} for the $\Delta$ resonance. However, for higher resonances ($N\neq0$) we use a slightly different form factor, similar to Ref.~\cite{Rein}, but with same assumptions in Ref.~\cite{GS}:
\begin{widetext}
\begin{eqnarray}\label{GS_FF}
F^V(W,k^2)=&& \frac{1}{2}\left( 1 - \frac{k^2}{(M+W)^2}\right)^{\frac{1}{2}}\left( 1 - \frac{k^2}{4M^2}\right)^{-\frac{N}{2}} \sqrt{3\left(G_V^{f_3}(W,k^2)\right)^2 + \left(G_V^{f_1}(W,k^2)\right)^2 }\nonumber\\
F^A(W,k^2)=&& \frac{\sqrt{3}}{2} \left( 1 - \frac{k^2}{(M+W)^2}\right)^{\frac{1}{2}}\left( 1 - \frac{k^2}{4M^2}\right)^{-\frac{N}{2}} \left[1 - \frac{W^2 + k^2 - M^2}{8M^2} \right] C_5^A(k^2),\nonumber\\
\end{eqnarray}
\end{widetext}
where $N$ is given in Table \ref{res_table} and $G_V^{f_3}(W,k^2)$ and $G_V^{f_1}(W,k^2)$ are given in Ref.~\cite{GS}.
$C_5^A(k^2)$ has a dipole form with two adjustable parameters, $M_A$ and $C_5^A(0)$, that can be fitted to data:
\begin{equation}\label{Axial_FF}
C_5^A(k^2)= \frac{C_5^A(0)}{\left(1 - \frac{k^2}{M_A^2} \right)^2}.
\end{equation}
\subsection{Nonresonant contribution}
\begin{figure}
  \centering
\includegraphics [width=0.9\linewidth]{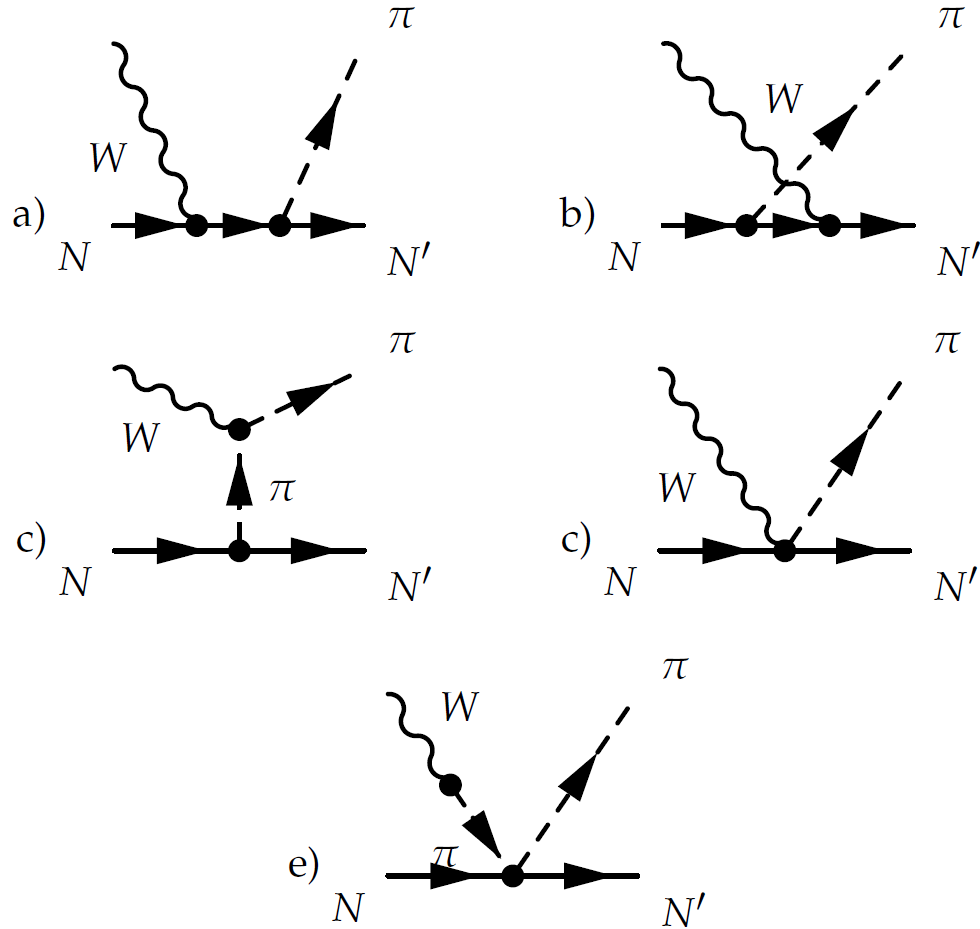}
\caption{Nonresonant pion production diagrams:  (a) nucleon pole (NP), (b) crossed nucleon pole (CNP),
(c) pion-in-flight (PIF), (d) contact term (CT), and (e) pion pole (PP)}
\label{HNV_diag}
\end{figure}

Nonresonant interactions are defined by a set of Feynman diagrams as shown in Fig.~\ref{HNV_diag}. The pseudovector $NN\pi$ vertices are
determined by the HNV model \cite{HNV}. It is an effective chiral field theory
based on the nonlinear $\sigma$ model \cite{GellMann}.\\
The corresponding amplitudes are
\begin{widetext}
\begin{align}
\mathcal{M}^{NP}_{CC} &= C^{NP}\cos\theta_C\frac{g_A}{\sqrt{2}f_{\pi}}~ \frac{1}{s-M^2}~ \bar{u}(p_2)~ \not{q}~\gamma_5 (\not{p_1} + \not{k} +M) \epsilon^{\mu} [F^{V}_{\mu} - F^A_{\mu}]~ u(p_1),\nonumber\\
\mathcal{M}^{CNP}_{CC}&= C^{CNP}\cos\theta_C\frac{g_A}{\sqrt{2}f_{\pi}}~\frac{1}{u-M^2}~ \bar{u}(p_2) ~\epsilon^{\mu}[F^V_{\mu}- F^A_{\mu}]~(\not{p_2} - \not{k} +M) \not{q}\gamma_5 ~u(p_1),\nonumber\\
\mathcal{M}^{PF}_{CC}&=  C^{PF}\cos\theta_C\frac{g_A}{\sqrt{2}f_{\pi}}~ \frac{1}{t-m_{\pi}^2}~ F_{PF}(k^2) \bar{u}(p_2)~\gamma_5~[2q\epsilon- k\epsilon]2M ~u(p_1),\nonumber\\
\mathcal{M}^{CT}_{CC}&=  C^{CT}\cos\theta_C\frac{1}{\sqrt{2}f_{\pi}} ~  \bar{u}(p_2)~\epsilon^{\mu}\gamma_{\mu}~[g_A F^V_{CT}(k^2)~\gamma_5 - F_{\rho}((k-q)^2)] ~u(p_1),\nonumber\\
\mathcal{M}^{PP}_{CC}&=  C^{PP}\cos\theta_C \frac{1}{\sqrt{2}f_{\pi}} ~ \bar{u}(p_2)~ \frac{\epsilon k}{k^2 - m^2_{\pi}}~F_{\rho}((k-q)^2)   \not{k} ~u(p_1),
\label{HNV}\end{align}
\end{widetext}
where
\begin{align}\label{VA_FF}
 (F^V)^{\mu} &= 2\left[F_1^V (k^2) \gamma^{\mu} - \mu_V\frac{F_2^V(k^2)}{2M}[\gamma^{\mu}, \not{k}]\right], \nonumber\\
 (F^A)^{\mu} &= - G_A(k^2) \left[\gamma^{\mu} \gamma_5 + \frac{\not {k}}{m_{\pi}^2 - k^2}k^{\mu}\gamma_5\right].\nonumber\\
\end{align}
The vector form factors are
\begin{eqnarray}
F_1^V(k^2) &= \frac{1}{2}\left( F_2^p(k^2) - F_1^n(k^2) \right),\nonumber\\
\mu_V F_2^V(k^2) &= \frac{1}{2} \left( \mu_p F_2^p(k^2) - \mu_n F^n_2(k^2) \right).
\end{eqnarray}
Similar to the HNV model \cite{HNV}, the parametrization of Galster {\it et al.} \cite{Gal} is used.
 The axial form factor for nonresonant interactions is:
\begin{eqnarray}
G_A(k^2) = \frac{g_A}{(1-k^2/M_A^2)^2} , ~~~~~~g_A=1.26,
\end{eqnarray}
where for this work $M_V= 0.84$ GeV and $M_A =1.05$ GeV. In addition,
\begin{align}\label{}
F_{\rho}(t) = 1/(1-t/M_{\rho}^2),
\end{align}
where $m_{\rho}= 0.7758$ GeV, as proposed in Ref.~\cite{HNV}.
Conservation of vector current (CVC) requires that:
\begin{eqnarray}\label{sameF}
F_{PF}(k^2) = F^V_{CT}(k^2)= 2F_1^V(k^2).
\end{eqnarray}
Isospin coefficients $C^{NP}, C^{CNP}, C^{PF}, C^{CT}$, and $C^{PP}$ are given in Table~\ref{isospin_bkg} for different neutrino and antineutrino channels.\\
To calculate the helicity amplitudes of the above diagrams at Eq.~(\ref{HNV}), first the invariant amplitudes ($V_k$ and $A_k$) need to be calculated from transition amplitudes,
\begin{align}\label{first_ex}
\langle~N\pi|e^{\rho}J_{\rho}|N ~\rangle = &\sum_k \bar{u}_N(p_{2}) [V_k(s,t,u)O(V_k) \nonumber\\
                                                                     - &A_k(s,t,u)O(A_k)] u_N(p_{1}),
\end{align}
 for each channel. The vector and axial vector invariant amplitudes for two CC channels are given in Table \ref{inv_amp}. Isospin symmetry allows us to find $V_k^{p\pi^0}$ ($A_k^{p\pi^0}$) in terms of $V_k^{p\pi^+}$ ($A_k^{p\pi^+}$) and $V_k^{n\pi^+}$($A_k^{n\pi^+}$):
\begin{align}\label{}
V_k^{p\pi^0} =& -\frac{1}{\sqrt{2}}\left[ V_k^{p\pi^+} - V_k^{n\pi^+} \right]\nonumber\\
A_k^{p\pi^0} =& -\frac{1}{\sqrt{2}}\left[ A_k^{p\pi^+} - A_k^{n\pi^+} \right]
\end{align}
\begin{table}
\centering
\caption{Isospin coefficients for neutrino and anti-neutrino channels. } \label{isospin_bkg}
\renewcommand{\arraystretch}{1.5}
\begin{ruledtabular}
 \begin{tabular}{|l|c c c |}
 {\footnotesize CC Channels}  & {\footnotesize$C_{NP}$} &{\footnotesize$C_{CNP}$}&{\footnotesize$C_{PF}=C_{CT}=C_{PP}$}   \\ [0.7ex]
 \hline
{\footnotesize$\nu p \rightarrow l^- p \pi^{+}$ } & $0$  & $1$ &  $1$  \\
{\footnotesize$\nu n \rightarrow l^- p \pi^{0}$ } &$\frac{1}{\sqrt{2}}$ & $-\frac{1}{\sqrt{2}}$ &  $-\sqrt{2}$  \\
{\footnotesize$\nu n \rightarrow l^- n \pi^{+}$ } & $1$  & $0$ &  $-1$  \\
\hline
{\footnotesize$\bar{\nu} n \rightarrow l^+ n \pi^-$} &  $0$  & $1$ &  $1$  \\
{\footnotesize$\bar{\nu} p \rightarrow l^+ n \pi^0$} & $-\frac{1}{\sqrt{2}}$ & $ \frac{1}{\sqrt{2}}$ &  $\sqrt{2}$ \\
{\footnotesize$\bar{\nu} p \rightarrow l^+ p \pi^-$} & $1$  & $0$ &  $-1$ \\
\end{tabular}
\end{ruledtabular}
\end{table}
Knowing invariant amplitudes allows for straightforward calculation of the isobaric amplitudes, $\mathscr{F}_k$ and $\mathscr{G}_k$, by using the required relations given in Appendix \ref{appA}. All helicity amplitudes for nonresonant CC interaction  in terms of $\mathscr{F}_k$ and $\mathscr{G}_k$ are given in Table~\ref{helicity_amp} of Appendix \ref{HAapp}.\\
Isospin symmetry also allows the calculation of helicity amplitudes for NC interactions from the CC helicity amplitudes \cite{HNV}. The helicity amplitudes for $e^+$ polarization in Table~\ref{helicity_amp} are zero since the outgoing lepton in NC interactions is neutrino. \\
\begin{table*}
\centering
\caption{Invariant amplitudes.}
\label{inv_amp}
\renewcommand{\arraystretch}{1.3}
\begin{ruledtabular}
 \begin{tabular}{lcl}
 Amplitude ~~~~~~ $\nu_{\mu} + p \rightarrow  \mu p \pi^{+}$ & ~~~~~~$ \nu_{\mu} + n \rightarrow \mu n \pi^{+}$    \\ [3pt]
 \hline
 $\begin{aligned}
  &\scalebox{0.01}{~}\\
 V_1~~~~~~~~&\frac{g_A}{\sqrt{2}f_{\pi}} \left(\frac{4M}{s-M^2} F_1(k^2) + \frac{2\mu_V F_2(k^2)}{M}\right)\\
 V_2~~~~~~~~&\frac{g_A}{\sqrt{2}f_{\pi}} ~\frac{1}{qk}~\frac{4M}{u-M^2} F_1(k^2)\\
 V_3~~~~~~~~&\frac{g_A}{\sqrt{2}f_{\pi}} \left( \frac{4}{u-M^2} \mu_VF_2(k^2) \right)\\
 V_4~~~~~-&\frac{g_A}{\sqrt{2}f_{\pi}} \frac{4}{u-M^2} \mu_VF_2(k^2)\\
 V_5~~~~~~~~&\frac{g_A}{\sqrt{2}f_{\pi}}~\frac{1}{qk}~ \frac{2M}{t-m_{\pi}} F_{\pi}(k^2)\\[3pt]
 \end{aligned}$
 &
 $\begin{aligned}
  &\scalebox{0.01}{~}\\
 & \frac{g_A}{\sqrt{2}f_{\pi}} \left(\frac{4M}{s-M^2} F_1(k^2) + \frac{2\mu_V F_2(k^2)}{M}\right)\\
 & \frac{g_A}{\sqrt{2}f_{\pi}} ~\frac{1}{qk}~\frac{4M}{s-M^2} F_1(k^2)\\
 -&\frac{g_A}{\sqrt{2}f_{\pi}} ~\frac{4}{s-M^2} ~\mu_V F_2(k^2)\\
 -&\frac{g_A}{\sqrt{2}f_{\pi}} ~\frac{4}{s-M^2}~\mu_V F_2(k^2)\\
 -&\frac{g_A}{\sqrt{2}f_{\pi}}~\frac{4M}{qk}~ \frac{1}{t- m_{\pi}^2} F_1(k^2)\\[3pt]
 \end{aligned}$
 \\\hline
 $\begin{aligned}
  &\scalebox{0.01}{~}\\
A_1~~~~~~~~~~& \frac{g_A}{\sqrt{2}f_{\pi}} ~\frac{2M}{u-M^2} G_A(k^2) \nonumber\\
A_3~~~~~~~~ -& \frac{g_A}{\sqrt{2}f_{\pi}} \left( \frac{2M}{u-M^2} G_A(k^2) \right)\nonumber\\
A_4~~~~~~~~ -&\frac{g_A}{\sqrt{2}f_{\pi}} ~\frac{1}{M} G_A(k^2) + \frac{1}{\sqrt{2}Mf_{\pi}} F_{\rho}\left((k-q)^2\right)\nonumber\\
A_7~~~~~~~~ -&\frac{g_A}{\sqrt{2}f_{\pi}}~ \frac{2M}{m^2_{\pi}-k^2} G_A(k^2)\nonumber\\
A_8~~~~~~~~ +&\frac{g_A}{\sqrt{2}f_{\pi}}~\frac{1}{m^2_{\pi}-k^2} \left( 1 + \frac{4M^2}{u-M^2} \right)G_A(k^2) \nonumber\\
~~~~~~~~ -& \frac{1}{\sqrt{2}f_{\pi}}~\frac{1}{m^2_{\pi}-k^2}F_{\rho}\left((k-q)^2\right)\\[3pt]
\end{aligned}$
&
$\begin{aligned}
 &\scalebox{0.01}{~}\\
-&\frac{g_A}{\sqrt{2}f_{\pi}} ~\frac{2M}{s-M^2} G_A(k^2)\nonumber\\
-&\frac{g_A}{\sqrt{2}f_{\pi}} ~\frac{2M}{s-M^2} G_A(k^2)\nonumber\\
 &\frac{g_A}{\sqrt{2}f_{\pi}} ~\frac{1}{M} G_A(k^2) - \frac{1}{\sqrt{2}f_{\pi}}\frac{1}{M} F_{\rho}\left((k-q)^2\right)\nonumber\\
-&\frac{g_A}{\sqrt{2}f_{\pi}}~ \frac{2M}{m^2_{\pi}-k^2} G_A(k^2)\nonumber\\
-&\frac{g_A}{\sqrt{2}f_{\pi}}~\frac{1}{m^2_{\pi}-k^2} \left( 1 + \frac{4M^2}{s-M^2} \right)G_A(k^2)\\[3pt]
+& \frac{1}{\sqrt{2}f_{\pi}}~\frac{1}{m^2_{\pi}-k^2}F_{\rho}\left((k-q)^2\right)\\[3pt]
\end{aligned}$
\\
\end{tabular}
\end{ruledtabular}
\end{table*}
\section{Results and Comparison with experiments}\label{result}
The model described in this work includes a full kinematic description of the final state particles for CC and NC (anti) neutrino-nucleon interactions. It has been calculated in the helicity basis which is very suitable for implementation in event generators. \\
The full model includes resonant and nonresonant interactions, as well as interference effects.
The resonance part of the model (which is based on the RS-model) includes $17$ resonances up to $M_R=2 GeV$ (see Table~\ref{res_table}), and is therefore valid up to $W=2~\text{GeV}$. For nonresonant interactions, the model is based on chiral symmetry and it is not reliable at high W. A practical solution \cite{Gil} for a complete model with resonant and nonresonant interactions, is to multiply a form factor\footnote{The proposed form factor in this work is:
 \begin{equation}\label{}
   F_{vir}(W) =
  \begin{cases}
    1       & \quad \text{if } W<1.4~\text{GeV}\\
    -23.31W^2 + 64.92W -44.2  & \quad \text{if } 1.4\text{ GeV}\leq W \leq 1.6\text{ GeV}\\
    0       & \quad \text{if } W>1.6~\text{GeV}\\
  \end{cases}\nonumber
\end{equation}}
 by the virtual pion propagator of the PIF diagram in Fig.~\ref{HNV_diag}.
CVC requires the inclusion of this form factor for several other amplitudes. This will reduce the nonresonant contributions smoothly in the $1.4~\text{GeV}\leq W\leq 1.6~\text{GeV}$ region, therefore, the nonresonant interaction will have no effect at $W\geq 1.6$ GeV. \\
The dipole form factor in Eq.~(\ref{Axial_FF}) is a function of $Q^2$. Therefore it is suitable to fit the adjustable parameters to differential cross section measurements in $Q^2$.
 The ANL experiment provided a measurement for the $\nu_{\mu} p \rightarrow \mu^- p \pi^+$ channel with the selections $0.5~\text{GeV}<E<6~\text{GeV}$ and $W<1.4~\text{GeV}$ \cite{ANL}.
 The best-fit values for the parameters can be found from a $\chi^2$ minimization fit to averaged $d\sigma/dQ^2$ over the ANL
   \begin{figure}
\centering
\includegraphics [width=\linewidth]{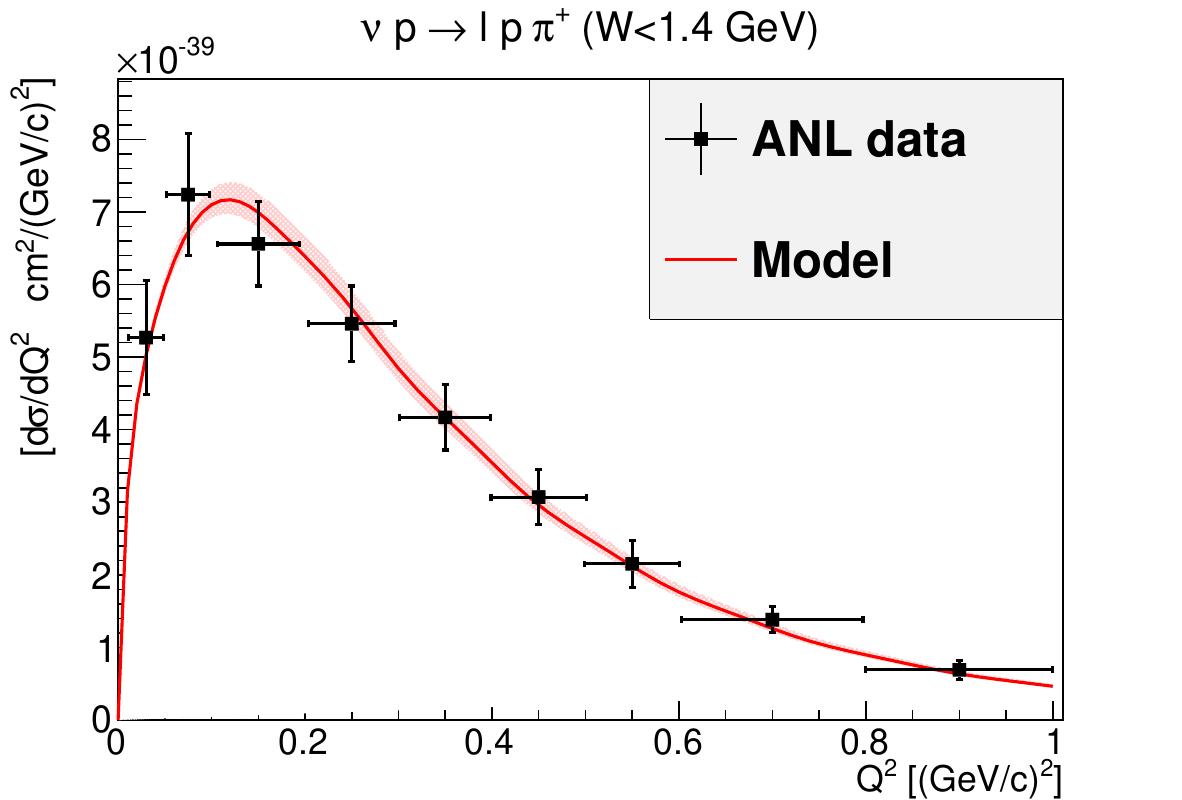}
\caption{$Q^2$-differential cross section ANL data with invariant mass cut, $W<1.4~\text{GeV}$. The prediction of the model with fitted parameters is shown with a solid red curve. The shaded area accounts for the variation of the results when $M_A$ changes within its error interval.}\label{ds_dQ}
\end{figure}
flux \cite{ANL_flux}. The results are
\begin{align}\label{}
  M_A = 0.733 \pm 0.068~\text{GeV},~~
  C_5^A= 0.993 \pm 0.101~\text{GeV}.
\end{align}
The Gaussian correlation coefficient, $r = −0.858$, shows that the parameters are strongly anticorrelated.
Figure \ref{ds_dQ} shows that the results of the fit with ANL data are within $1\sigma$ error bars.\\
 \begin{figure*}
\centering
 \begin{minipage}{.52\textwidth}
 \includegraphics[width=\textwidth]{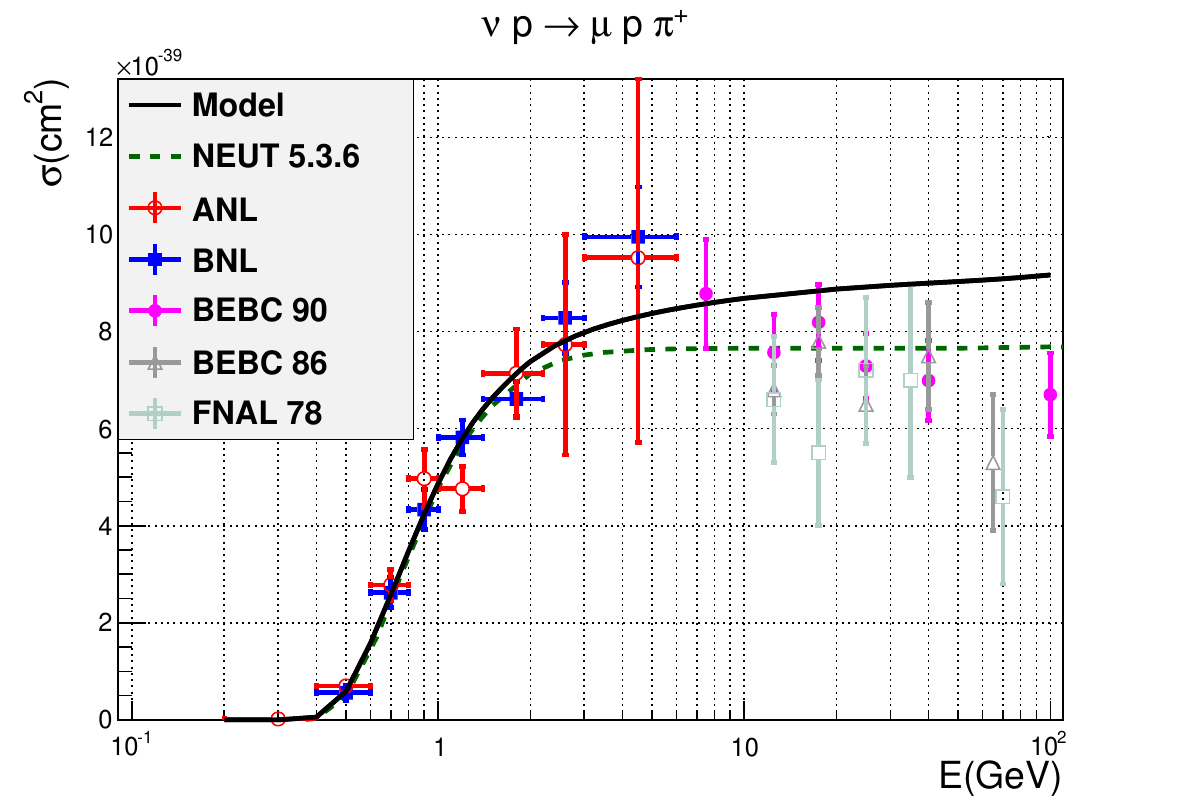}
\label{PP_high}
\end{minipage}
\begin{minipage}{.35\textwidth}
  \includegraphics[width=\textwidth]{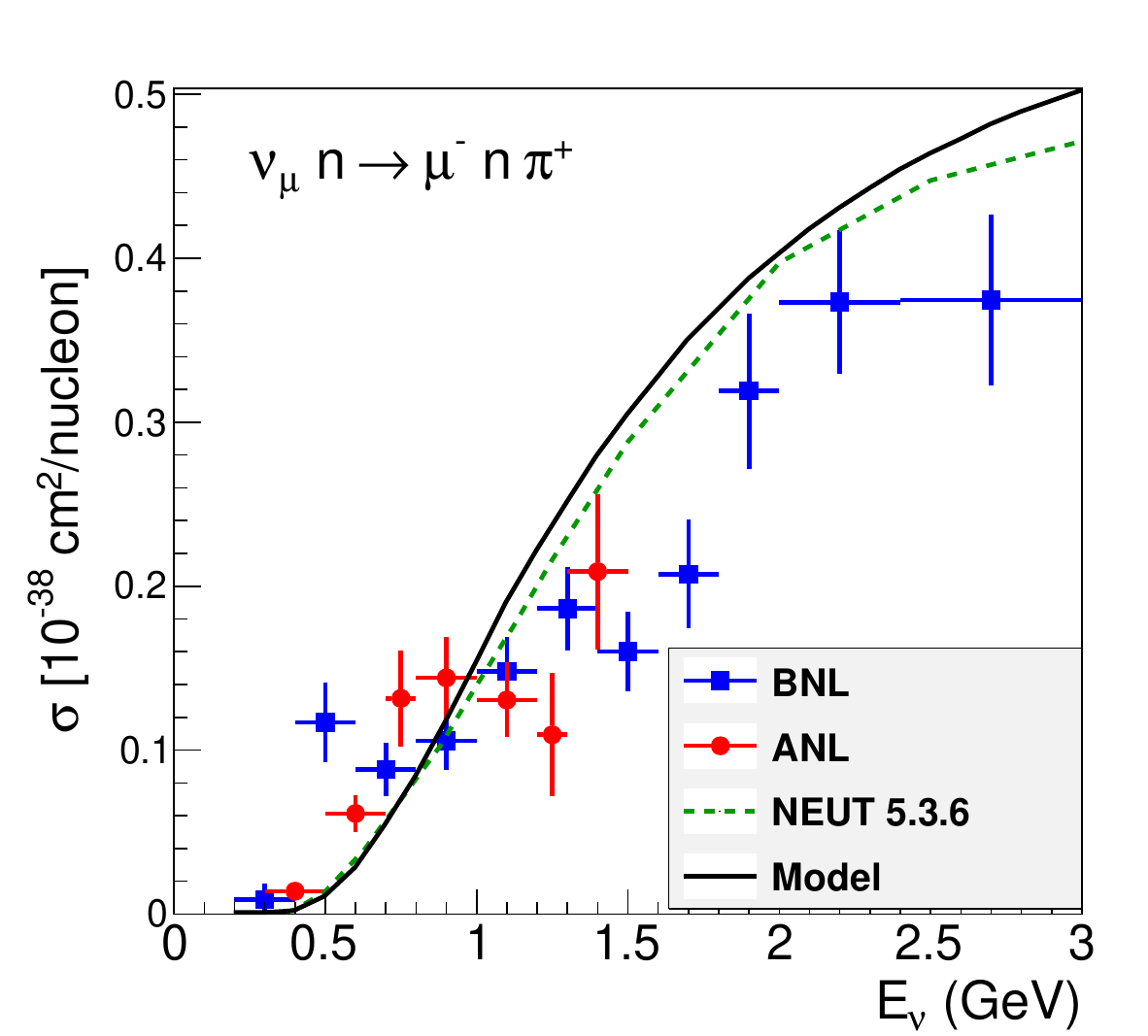}
\end{minipage}%
\begin{minipage}{.38\textwidth}
  \includegraphics[width=\textwidth]{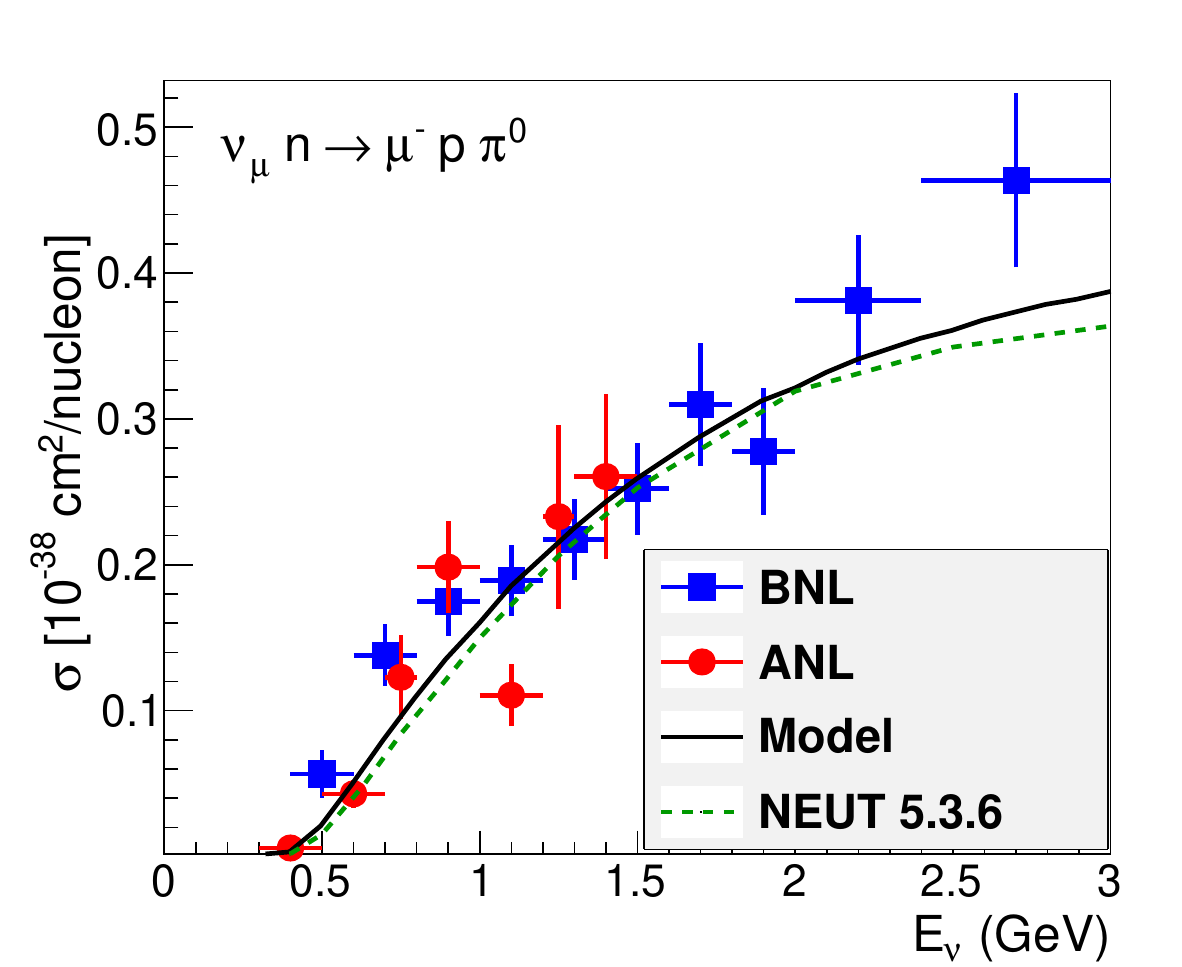}
\end{minipage}
  \caption{Total cross section for the $\nu p \rightarrow \mu p \pi^+$ (top), $\nu n \rightarrow \mu n \pi^+$(bottom left), and $\nu n \rightarrow \mu p \pi^0$ (bottom right) channels. Reanalyzed ANL and BNL data are from \cite{Wilkinson,Phil} and data from BEBC and FNAL, with an invariant mass cut $W<2~\text{GeV}$, are from \cite{BEBC, Barlag, FNAL78}. Curves are predicted cross sections by the model (solid black) and NEUT 5.3.6 (dashed green), with an invariant mass cut $W<2~\text{GeV}$.}
  \label{ABNL_re}
\end{figure*}
 For the rest of this section, we will show a comparison between the model predictions and bubble chamber CC and NC (anti)neutrino data. The RS-model is the default model for SPP in NEUT \cite{NEUT}; therefore, the NEUT predictions are also shown for comparison.\\
\subsection{Model and NEUT comparison with bubble chamber data}\label{NEUT_sub}
\begin{figure*}
  \centering
 \begin{minipage}{.46\textwidth}
    \includegraphics[width=\textwidth]{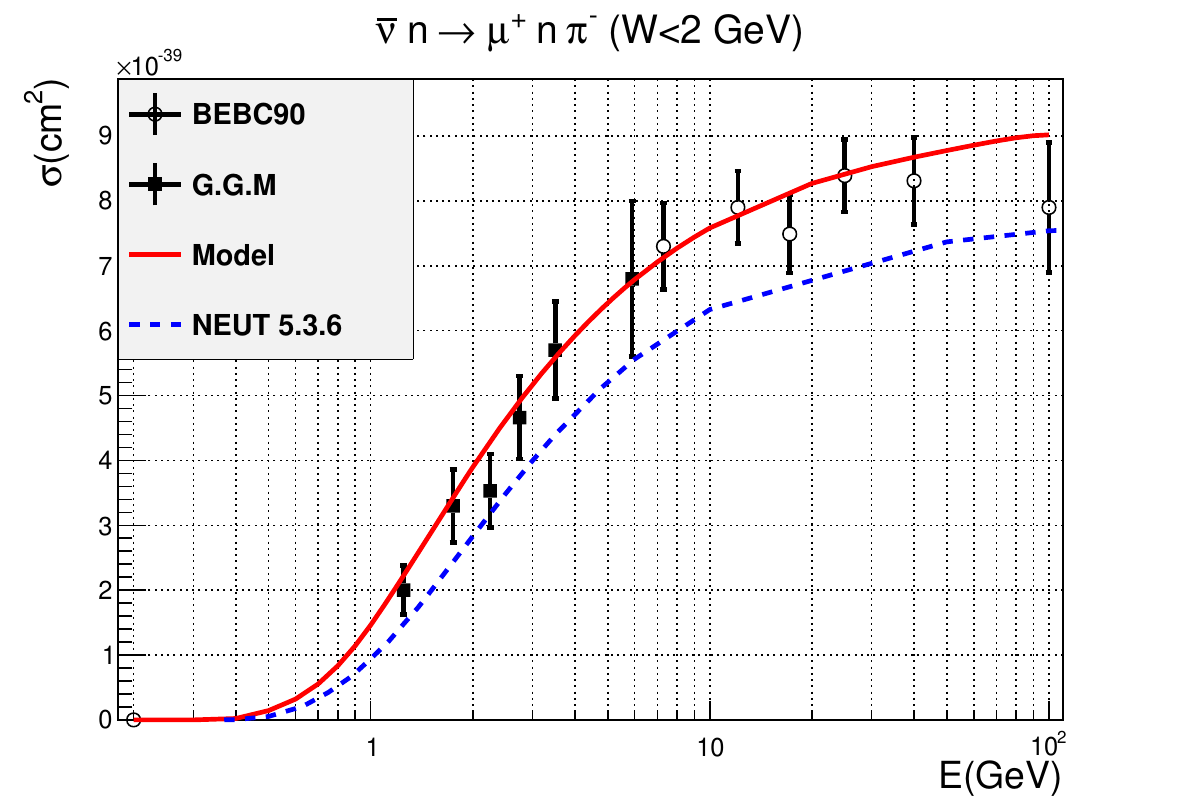}
  \end{minipage}
  \begin{minipage}{.46\textwidth}
    \includegraphics[width=\textwidth]{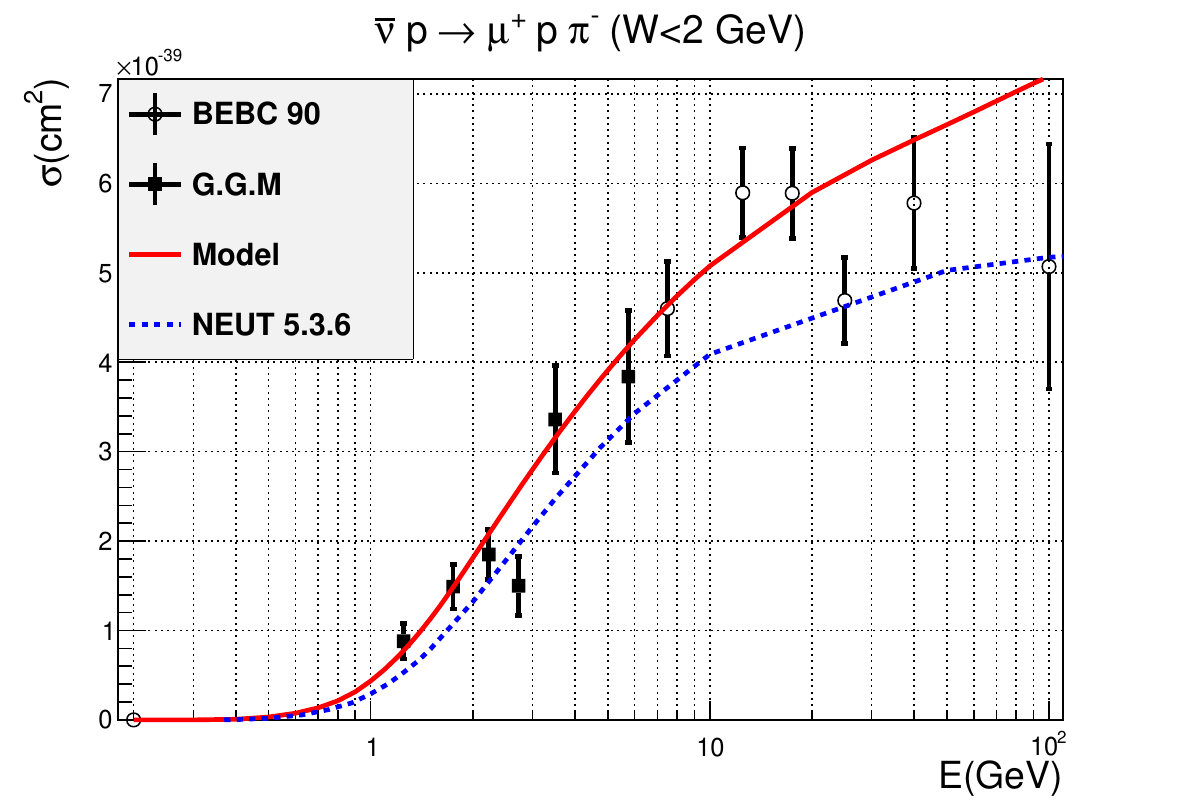}
  \end{minipage}
  \caption{Total cross section for two channels $\bar{\nu} n \rightarrow \mu^{+} n \pi^-$ (left) and $\bar{\nu} p \rightarrow \mu^{+} p \pi^-$ (right), as a function of neutrino energy. Data are from BEBC \cite{BEBC} and Gargamelle \cite{Barlag}, and curves are predicted cross sections by the model (solid black) and NEUT (dashed green) with an invariant mass cut $W<2~\text{GeV}$.}
  \label{CCanu_xsec}
\end{figure*}

\begin{figure}
\centering
\includegraphics [width=1.05\linewidth]{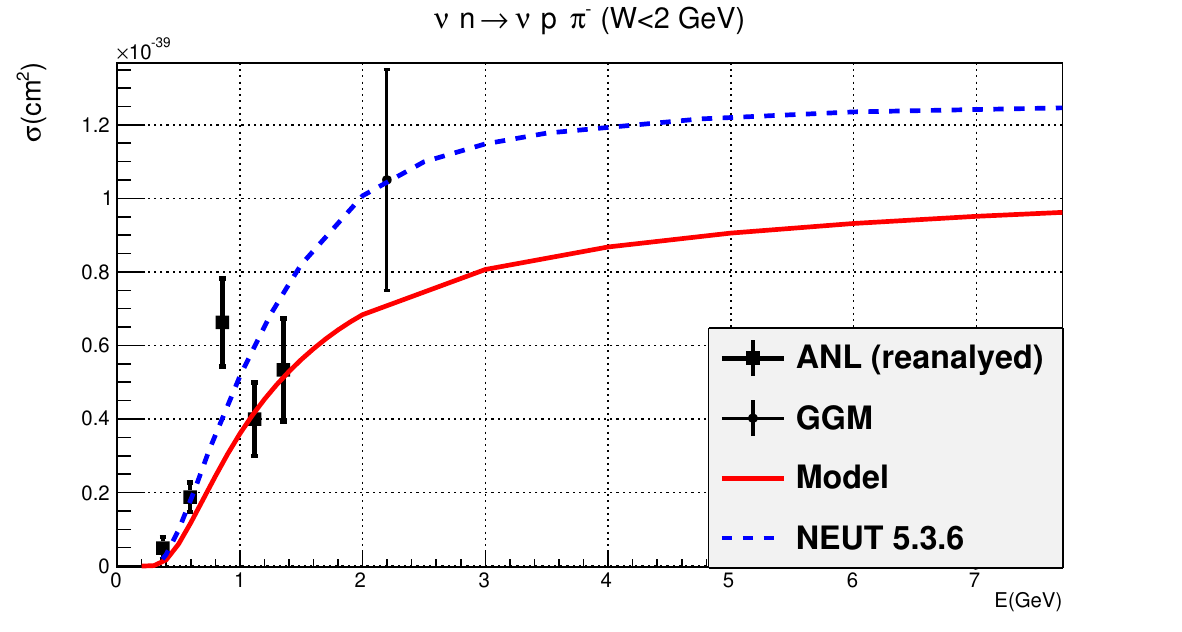}
\caption{Total cross section as a function of neutrino energy for the $\nu n \rightarrow \nu p \pi^{-}$ channel. The model (solid red) and NEUT predictions (dashed blue) have an invariant mass cut $W<2$ GeV. Data are from \cite{sam}} \label{NC_piminus}
\end{figure}

\begin{figure*}
  \centering
\begin{minipage}{.46\textwidth}
    \includegraphics[width=\textwidth]{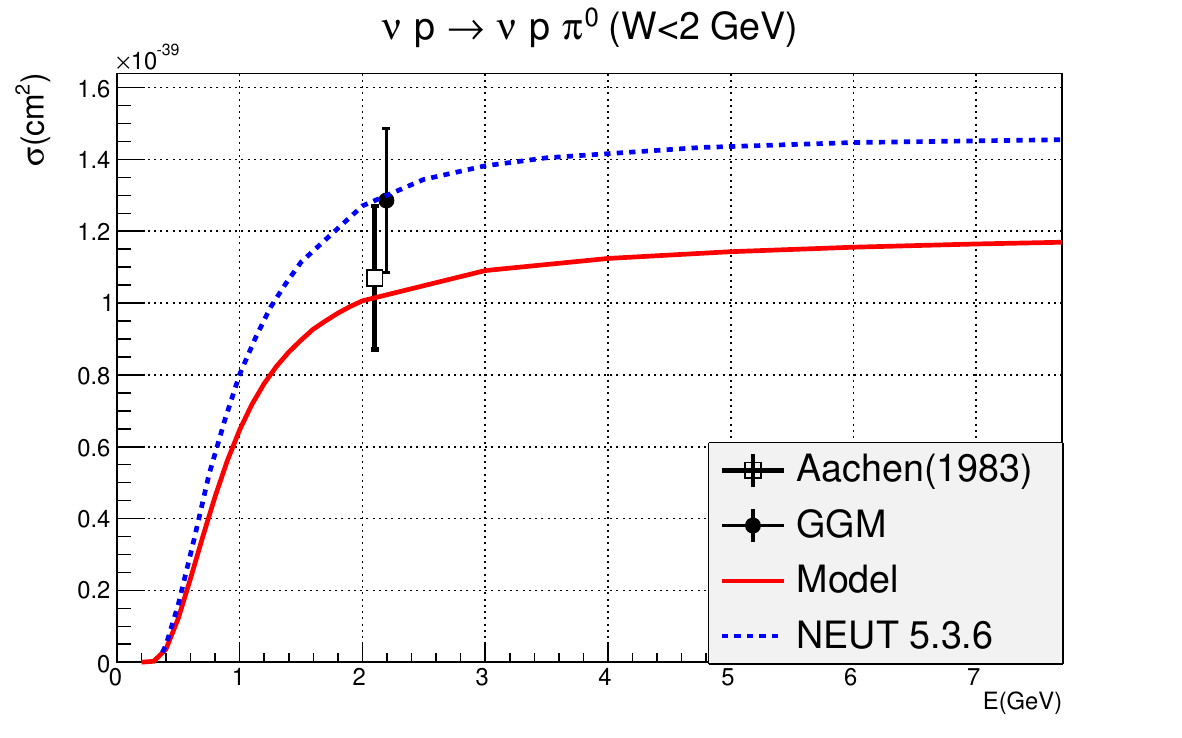}
  \end{minipage}
\begin{minipage}{.46\textwidth}
    \includegraphics[width=\textwidth]{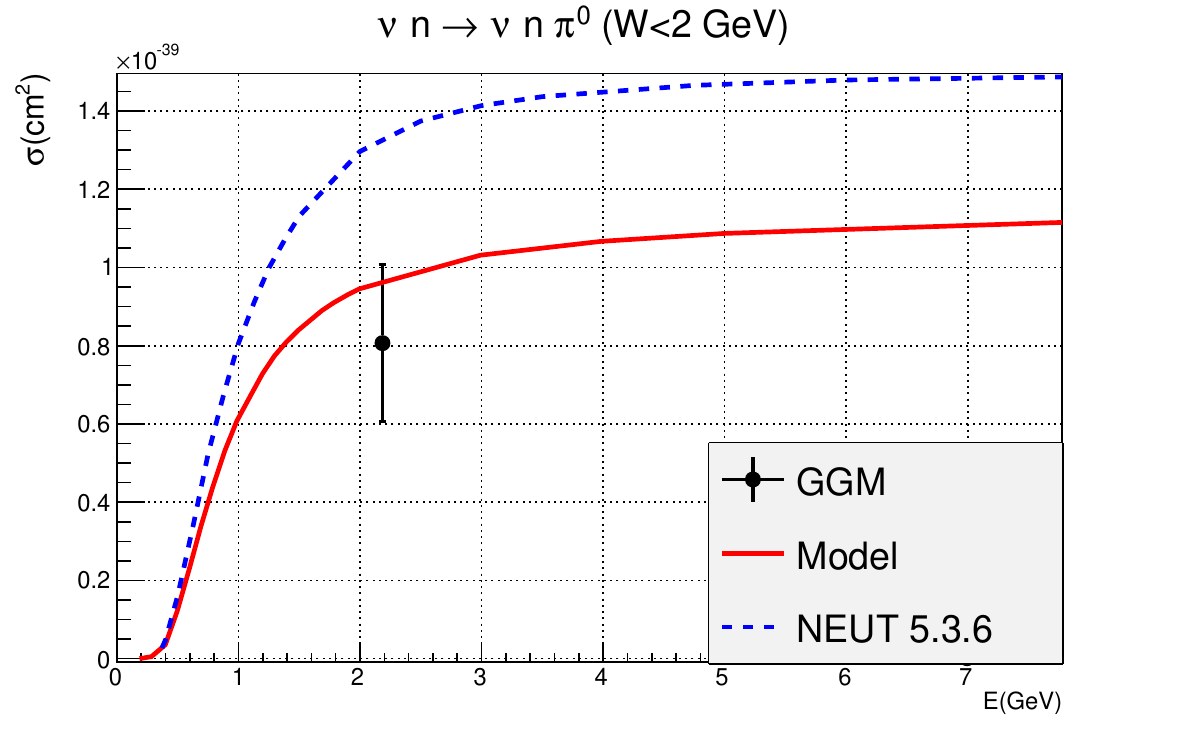}
  \end{minipage}
  \begin{minipage}[b]{0.46\textwidth}
    \includegraphics[width=\textwidth]{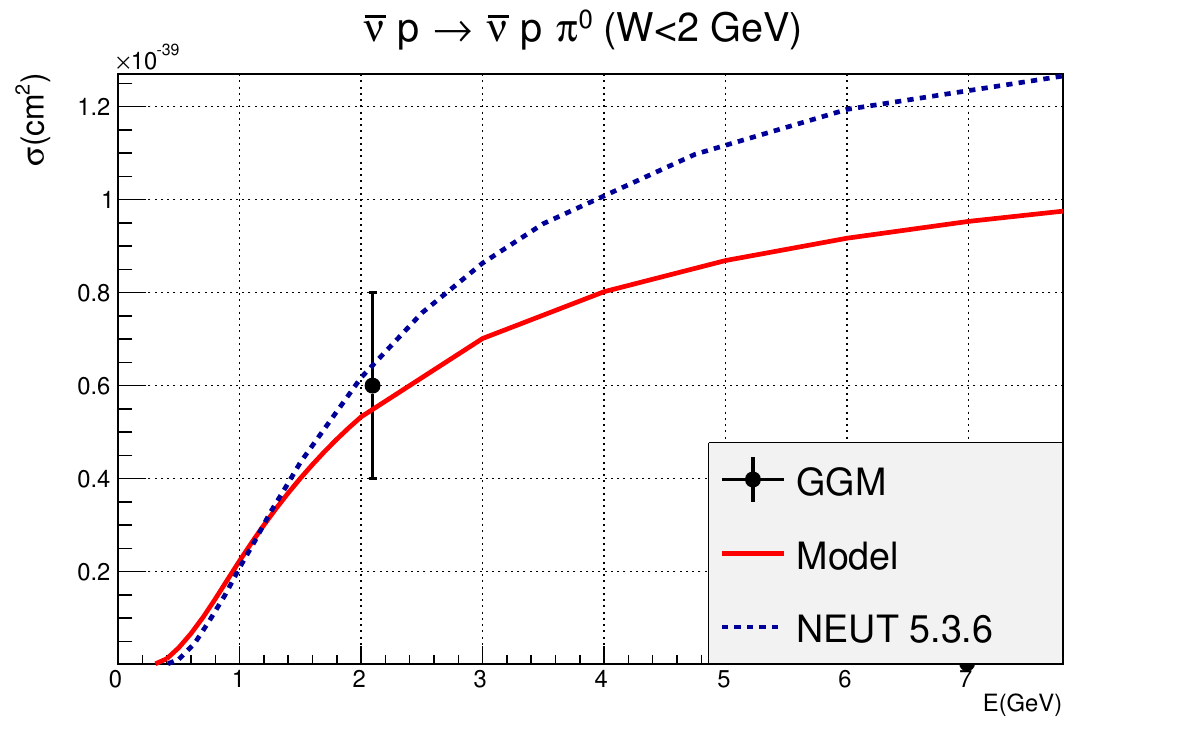}
  \end{minipage}
    \begin{minipage}[b]{0.46\textwidth}
    \includegraphics[width=\textwidth]{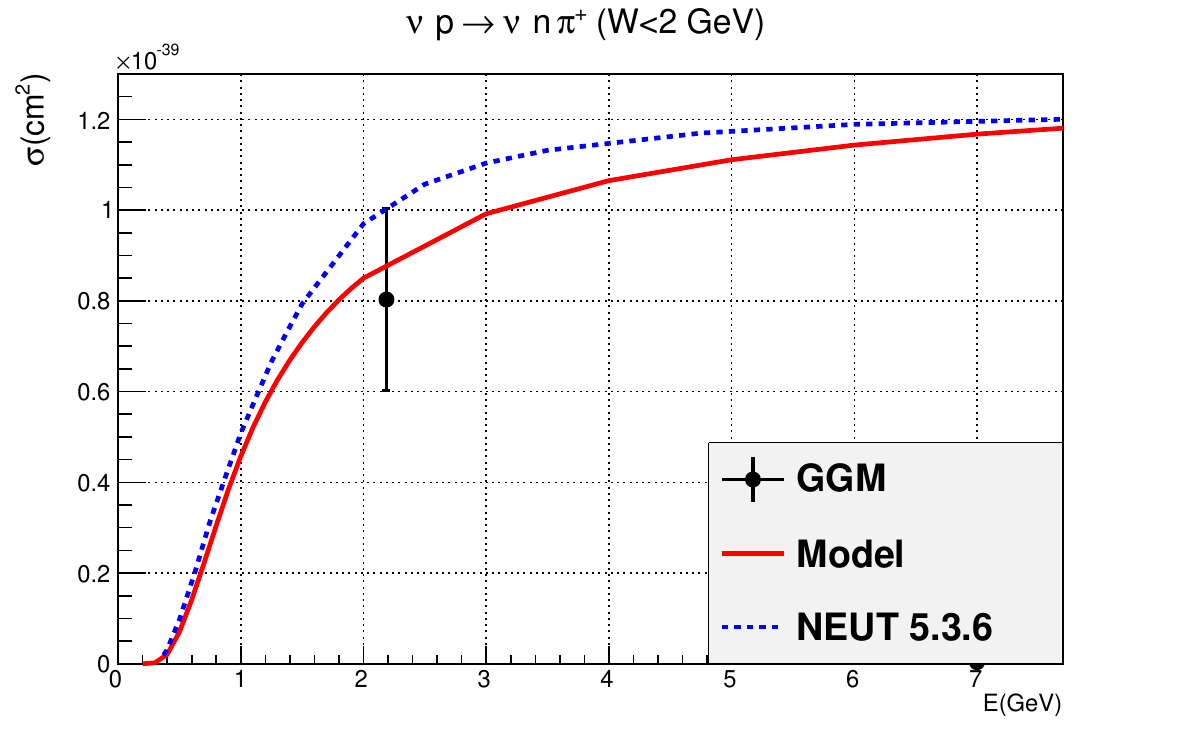}
  \end{minipage}
  \caption{Total cross section for NC (anti)neutrino channels. Data are from \cite{sam}, and curves are model(solid red) and NEUT (dashed blue) predictions with an invariant mass cut $W<2$ GeV.}
  \label{NC_xsec}
\end{figure*}

\begin{figure*}
  \centering
  \begin{minipage}{0.49\textwidth}
    \includegraphics[width=\textwidth]{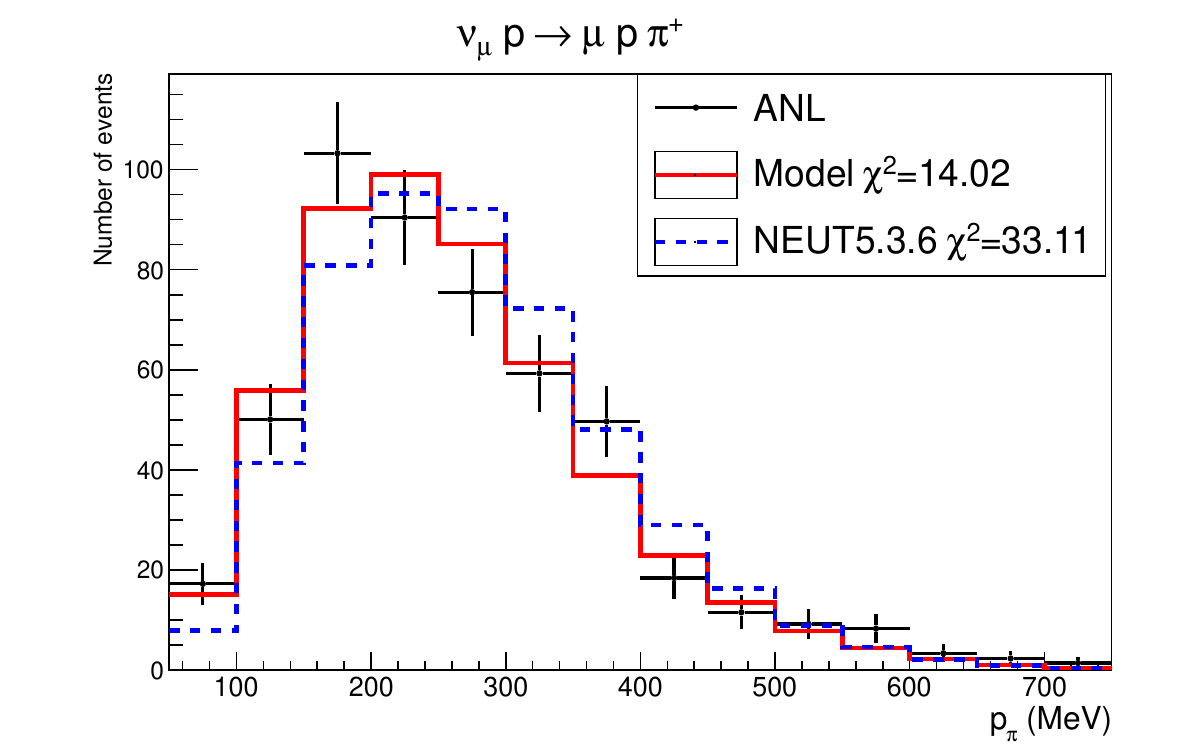}
  \end{minipage}
  \begin{minipage}{0.49\textwidth}
    \includegraphics[width=\textwidth]{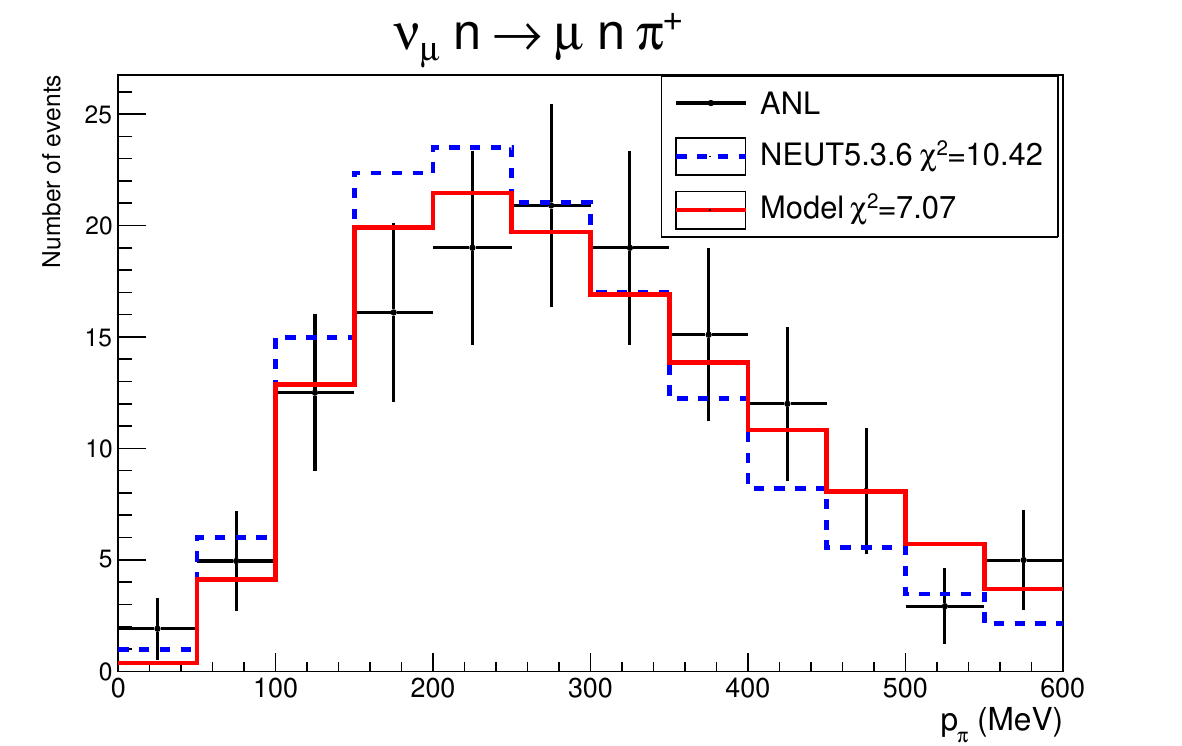}
  \end{minipage}
  \caption{Pion momentum distribution for the $\nu p \rightarrow \mu p \pi^+$ (left) and $\nu n \rightarrow \mu n \pi^+$ (right) channels from ANL \cite{ANL_pimom}.
  Model (solid red) and RS model (dashed blue) predictions of flux-averaged $p_{\pi}$-differential cross sections (with $W<2$ GeV cut), normalized to data are also presented for the two channels.}
  \label{pimom_dis}
\end{figure*}

In this section, the model defined in this paper and NEUT 5.3.6 are compared with bubble chamber data for SPP channels.
The SPP model in NEUT 5.3.6 is the RS-model with GS form factor \cite{GS}, including the isospin 1/2 background contribution with an adjustable
coefficient defined in the original paper \cite{RS}.\\
Several bubble-chamber experiments have measured the total cross section as a function of neutrino energy. The ANL \cite{ANL} and BNL \cite{BNL} experiments have measured the CC neutrino channels with a low energy neutrino beam. These data have been reanalyzed recently \cite{Wilkinson}.
Figure \ref{ABNL_re} shows the reanalyzed ANL and BNL data from \cite{Wilkinson,Phil}, as well as data from BEBC \cite{BEBC} and FNAL \cite{FNAL78} which utilize higher energy neutrino fluxes.
The model and NEUT predictions with an invariant mass cut of $W\leq2~\text{GeV}$ are also included for comparison.\\
For antineutrino, the BEBC experiment \cite{BEBC} on a deuterium target and the Gargamelle experiment \cite{GGM} on a propane target measured the total cross section for the $\bar{\nu} p \rightarrow \mu^{+} p \pi^-$ and $\bar{\nu} n \rightarrow \mu^{+} n \pi^-$ channels.
Figure \ref{CCanu_xsec} shows the data, model, and NEUT predictions with an invariant mass cut of $W<2~\text{GeV}$. Gargamelle data are normalized to the proton and neutron cross sections based on Ref.~\cite{Barlag}.
There are few available bubble-chamber data for NC SPP channels. These are from ANL \cite{ANL_NC} (deuterium target) and Gargamelle (propane). For the $\nu n \rightarrow \nu p \pi^{-}$ channel, the model and NEUT predictions are compared with Gargamelle and reanalyzed ANL data (based on \cite{Phil}) in Fig.~\ref{NC_piminus}.\\
There are also a few measurements for other NC neutrino and antineutrino channels, by Gargamelle \cite{GGM_NC} and the Aachen-Padova spark chamber \cite{Aachen}. The model and NEUT predictions are compared with all available data in Fig.~\ref{NC_xsec}.\\
ANL also measured the pion momentum distribution in the lab frame for two CC channels: $\nu_{\mu} + p \rightarrow  \mu p \pi^{+}$, and $\nu_{\mu} n\rightarrow \mu n \pi^{+}$ \cite{ANL_pimom}. To compare the model to predictions in the lab frame one needs to generate events in the isobaric frame and boost it to the lab frame. This is done with an implementation of the model in NEUT \cite{NEUT} and the plots are made by NUISANCE \cite{NUISANCE} as shown in Fig.~\ref{pimom_dis}.\\
\subsection{$W$ distribution}
\begin{figure*}
\centering
\includegraphics [width=1.\linewidth]{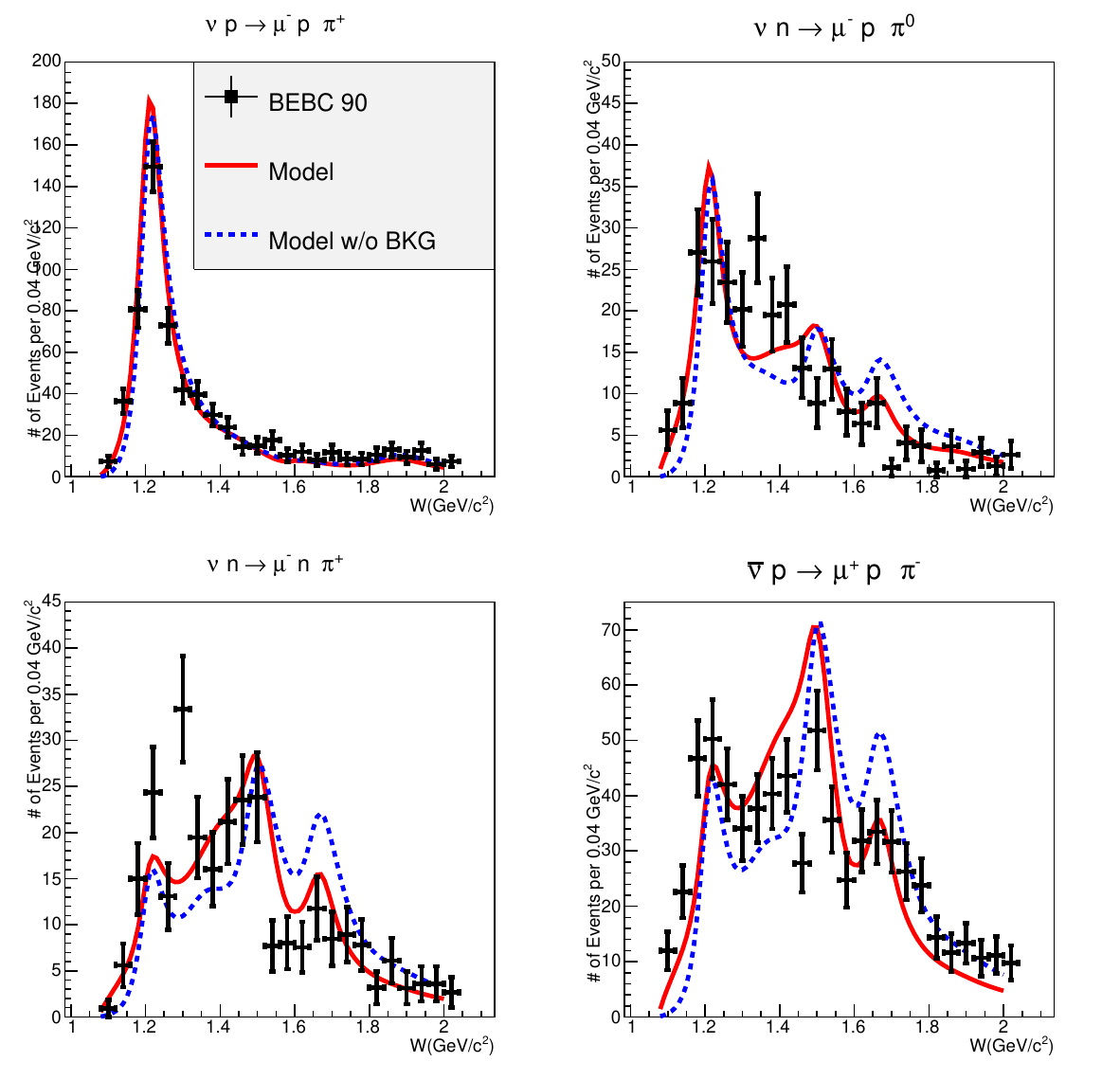}
\caption{W distribution for different neutrino and antineutrino CC channels from Ref.~\cite{BEBC}. Curves are the model predictions with (solid red) and without (dashed blue) background.}\label{BEBC_W}
\end{figure*}

\begin{figure*}
  \centering
  \begin{minipage}{0.47\textwidth}
    \includegraphics[width=\textwidth]{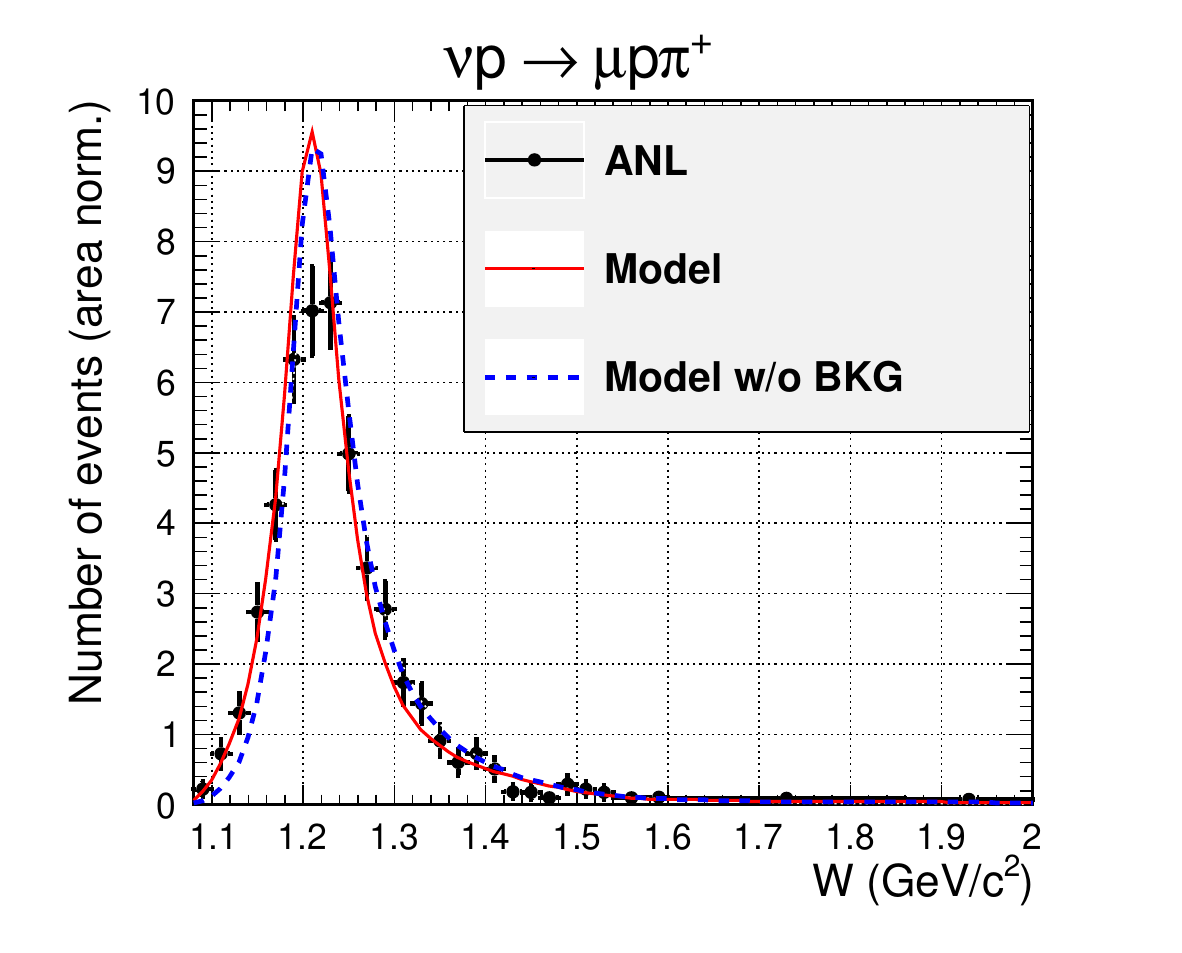}
  \end{minipage}
  \begin{minipage}{0.47\textwidth}
    \includegraphics[width=\textwidth]{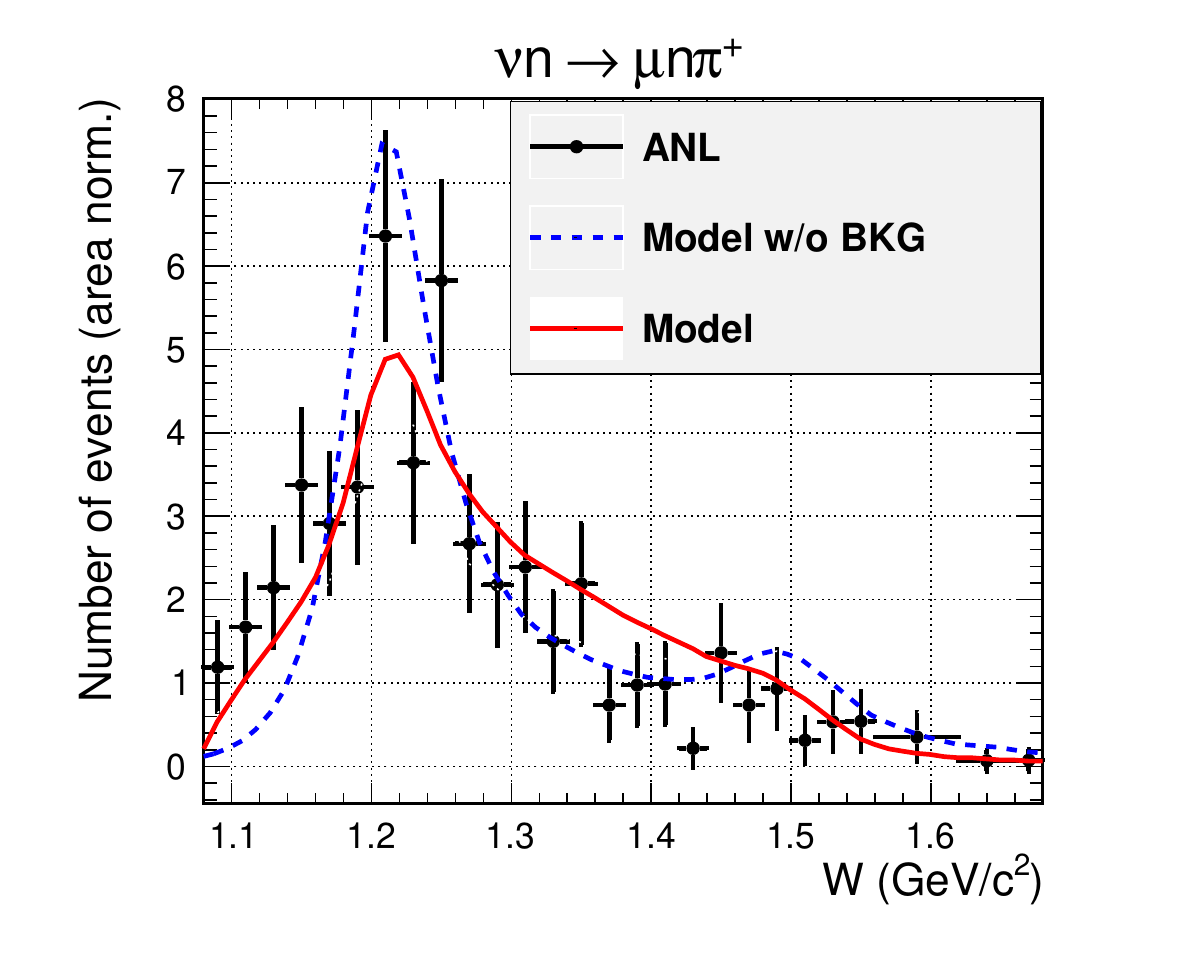}
  \end{minipage}
  \begin{minipage}{0.47\textwidth}
    \includegraphics[width=\textwidth]{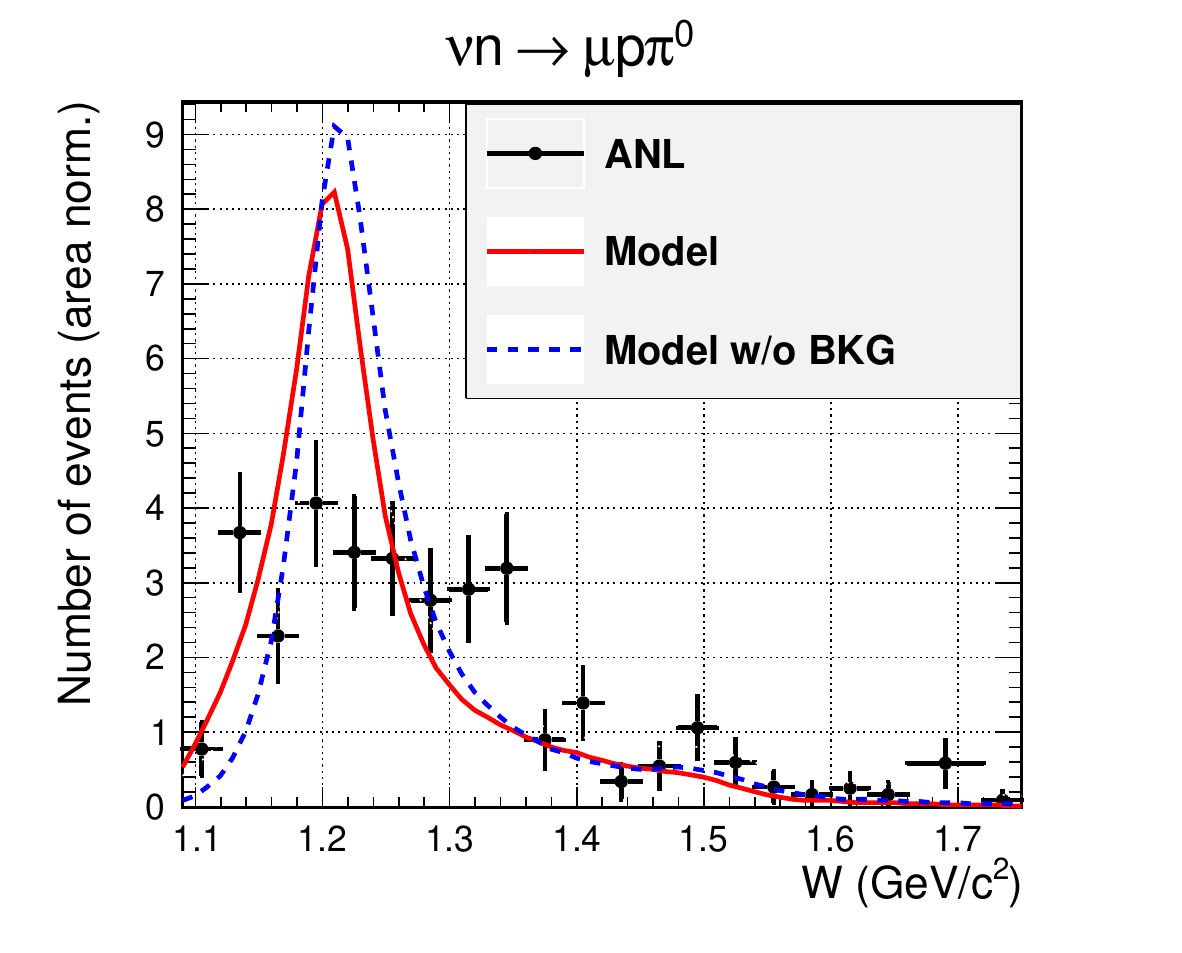}
  \end{minipage}
  \caption{W distribution for different neutrino CC channels from Ref.~\cite{ANL}. Curves are the model predictions with (solid red) and without (dashed blue) background.}
 \label{ANL_Wdis}
\end{figure*}

Distribution of the invariant mass of the hadronic system, $W$, provides information about individual resonance contributions where each resonance has a peak around its own resonance mass. The BEBC experiment measured the $W$ distribution with neutrino and antineutrino beams.
 The relatively high (anti) neutrino energy flux in this experiment showed clear patterns for the different channels.
 A shape comparison requires an area-normalized flux averaged over \cite{Barlag} $d\sigma/dW$.
 Figure \ref{BEBC_W} shows the model comparisons with BEBC data \cite{BEBC}. To demonstrate the effect of the nonresonant background, the model prediction without the nonresonant contribution is also shown for comparison.
 It is apparent that the $\nu p \rightarrow \mu p \pi^+$ channel, with isospin $3/2$ contributions, is dominated by $\Delta(1232)$ resonance production. Other channels are a combination of both isospin $1/2$ and $3/2$ resonances. Therefore, few bumps appear at higher $W$ due to the isospin $1/2$ resonances.\\
  At lower energy, the same comparison with ANL \cite{ANL} data is shown in Fig.~\ref{ANL_Wdis}. The model predictions with and without nonresonant background show the effects of the nonresonant contributions and its interference with resonances.
\subsection{Angular distribution}\label{angular_dis}
\begin{figure*}
  \centering
  \begin{minipage}{0.49\textwidth}
    \includegraphics[width=\textwidth]{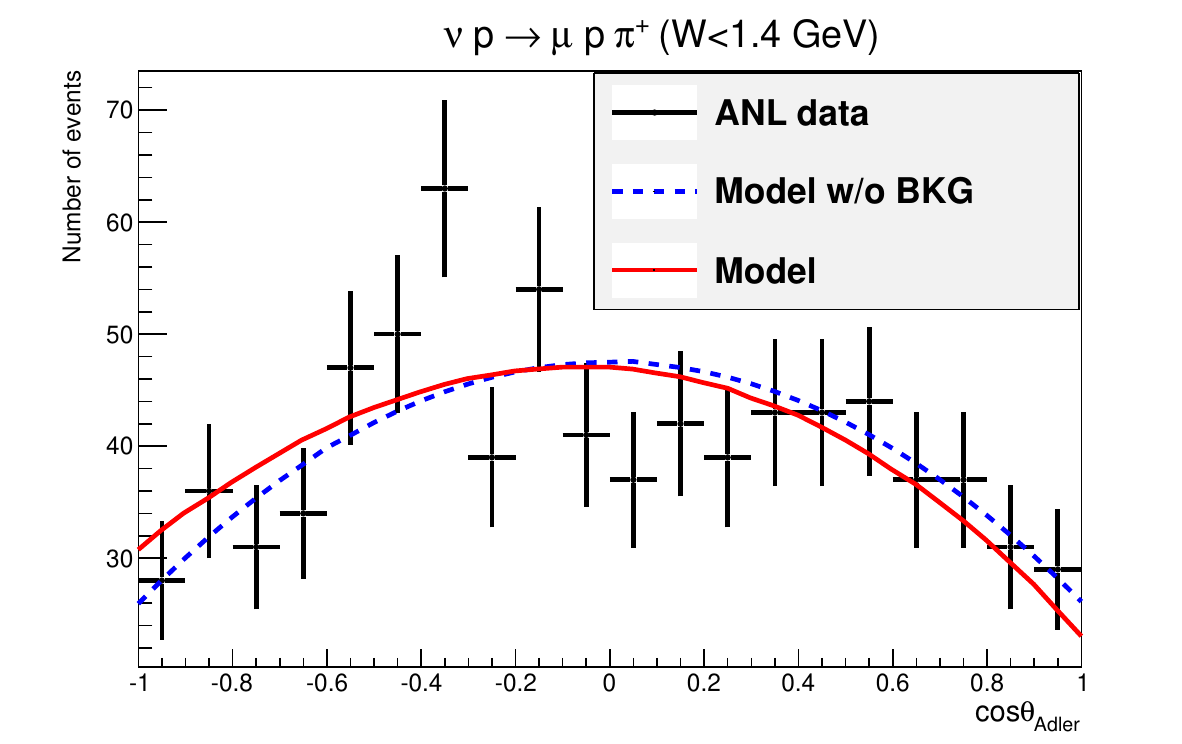}
  \end{minipage}
  \begin{minipage}{0.49\textwidth}
    \includegraphics[width=\textwidth]{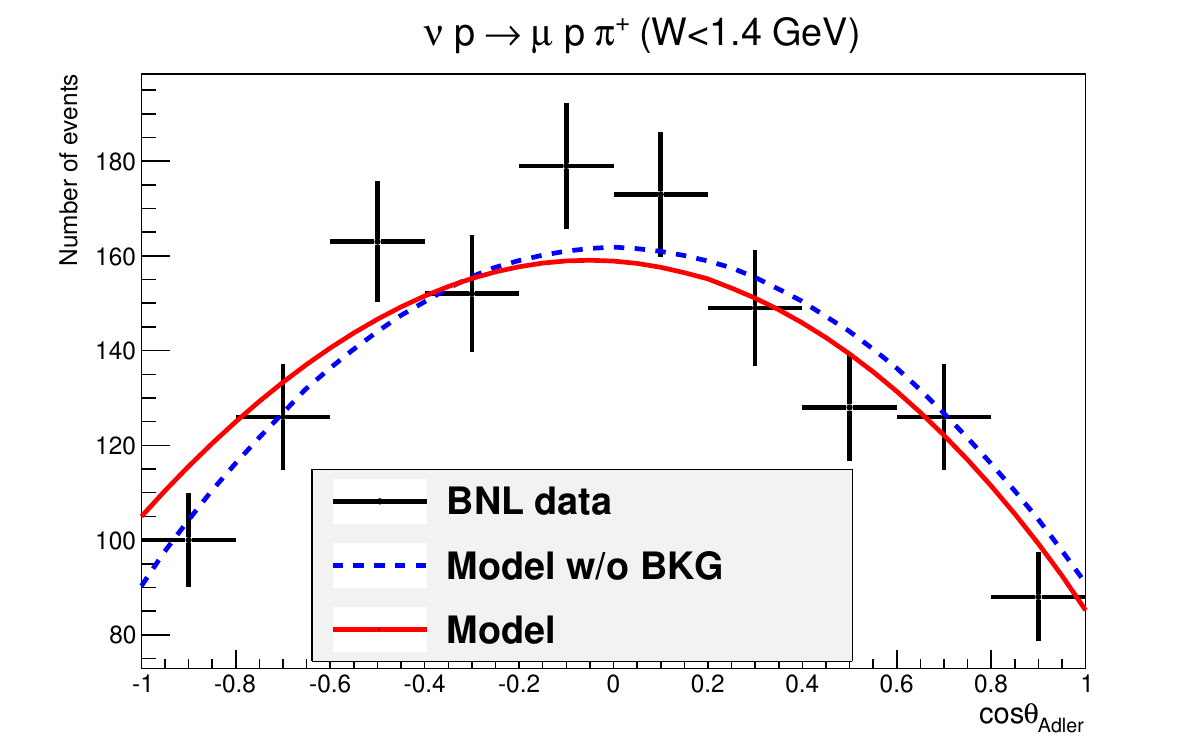}
  \end{minipage}
\caption{Event distribution in the pion polar angle for ANL (left) and BNL (right) with invariant mass cut, $W<1.4\text{ GeV}$, from Refs.~\cite{ANL, BNL}. Curves are flux-averaged, area-normalized prediction of the model (solid red) and the model without background (dashed blue).}
\label{ABNL_ang}
\end{figure*}

\begin{figure*}
\center
\includegraphics[width=1.\linewidth]{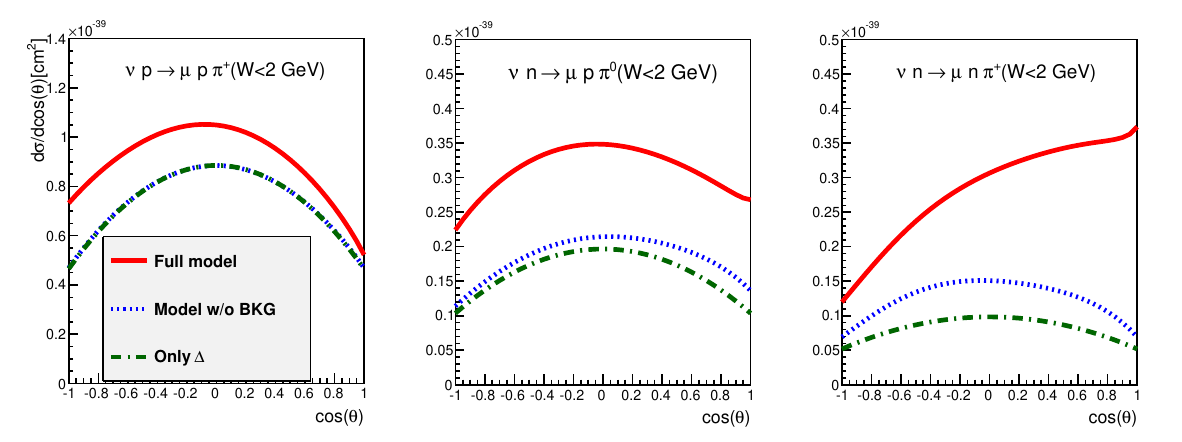}
\caption{\label{Polar}The differential cross section averaged over T2K flux in terms of the polar angle. The dotted blue curve shows the model prediction for resonant interaction (up to $W=2 GeV$), while the dashed green curve is only for the dominant $\Delta$ resonance. The solid red curve shows the full model.}
\end{figure*}

\begin{figure*}
\center
\includegraphics[width=1.\linewidth]{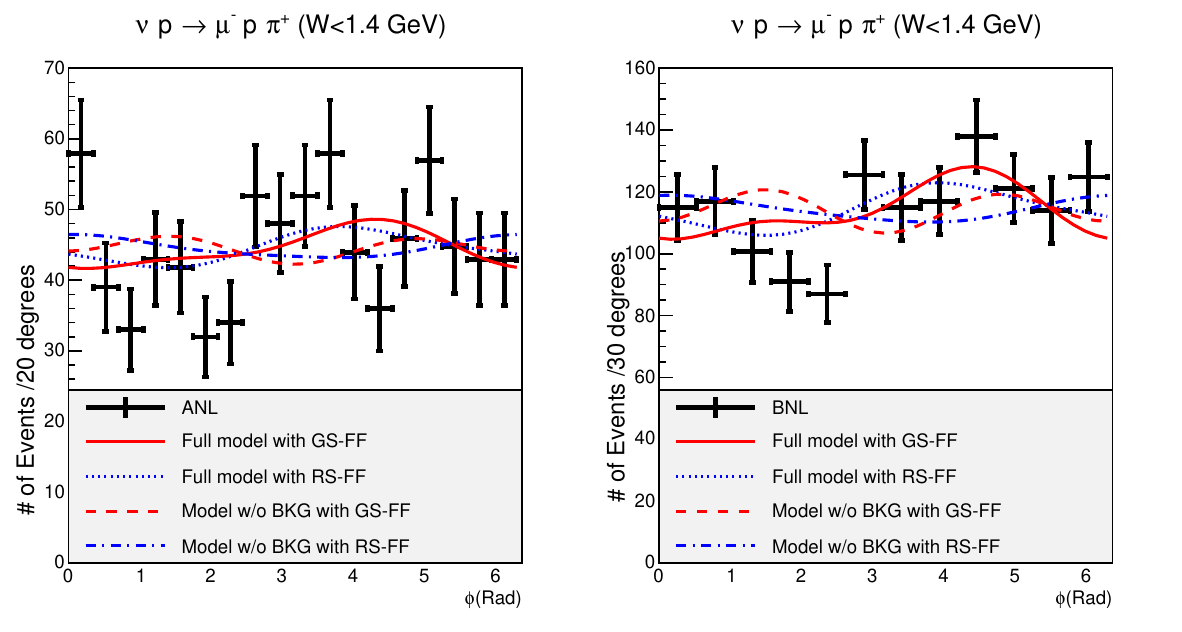}
\caption{ANL and BNL distribution of events in the pion azimuthal angle in the $\pi N$ rest frame
with $W < 1.4GeV$ for the $\mu^{-} p \pi^{+}$ final state. Curves are flux-averaged,
area-normalized prediction of the model for $d\sigma/d\phi$.}
\label{phi}\end{figure*}

 Polar ($\theta$) and azimuthal ($\phi$) angles are shown in the $N\pi$ rest frame in Fig.~\ref{Isoframe}. The $\theta$-distribution of an individual resonance is symmetric in the $N\pi$ rest frame; therefore, any modification from the symmetric pattern is caused by interference effects.
 The $\theta$-distribution for the $\nu p \rightarrow \mu p \pi^+$ channel has been measured by the ANL \cite{ANL} and the BNL \cite{BNL} experiments in the $\Delta$ region ($W<1.4\text{ GeV}$). The data are compared with the flux-averaged differential cross section predicted by the model in Fig.~\ref{ABNL_ang}. For these comparisons, the model was area normalized to the data.
 The symmetric prediction of the model without the nonresonant background contribution is also included for comparison.\\
 To show the effects of nonresonant interactions as well as interference with resonances, the full model (resonant and nonresonant up to $W=2\text{ GeV}$) and the resonance contribution of the model for CC neutrino channels are shown in Fig.~\ref{Polar}. The symmetric $\Delta$ contributions are also included for comparison. The differential cross section averaged over the T2K flux is shown for all these models.\\
 It is apparent from Fig.~\ref{Polar} that the nonresonant interference has a significant effect on the $\theta$ distribution (compare the solid red curves with the blue dotted curves). The interference between resonances has a non-negligible effect, especially on channels with isospin $1/2$. In the $\nu p \rightarrow \nu p \pi^{+}$ channel, only resonances with isospin $3/2$ can contribute and the $\Delta$ is dominant. Therefore, the effects of other resonances are negligible for this channel. \\
 In terms of pion angles, neutrino generators like NEUT \cite{NEUT} and GENIE \cite{GENIE} only have a contribution from the $\Delta$ resonance. They are missing all the other resonances and their interferences, as well as the nonresonance effects. Comparing shapes between this model and the $\Delta$ resonance contribution in Fig.~\ref{Polar} also shows the difference between the model and what is currently in generators.\\
The azimuthal angle ($\phi$) in the plane perpendicular to the momentum transfer (see Fig.~\ref{Isoframe}) is sensitive to interference effects. It is also a good observable to extract form factors.
 For the RS-model and resonant interactions, there are two available form factors: the dipole (RS) form factors from the original RS model \cite{RS}, and the GS form factors\footnote{It is called a GS form factor, but in fact, Eq.~(\ref{GS_FF}) is different from the GS form factors in Ref.~\cite{GS} for higher resonances.} introduced in Eq.~(\ref{GS_FF}).\\
Figure \ref{phi} shows ANL \cite{ANL} and BNL \cite{BNL} event distribution in $\phi$ with model predictions for two form factors which are notably different.
The model predictions without nonresonant background are also included, where they produce different shapes than the full model.\\
According to \cite{Federico},
 the shape of the $\phi$ distribution is almost unaffected by nuclear effects. Therefore experiments with a nuclear target are sensitive to the axial form factors, while bubble
chamber data is not precise enough for this purpose.
\subsection{Conclusion}
The model proposed in this work provides a differential cross section, $d\sigma/dWdQ^2d\Omega$, for single pion production up to $W=2~\text{GeV}$. It consists of resonant and nonresonant interactions and includes interference effects.\\
 Bubble-chamber data are used to extract the axial form factor of the resonant contributions. The model has good agreement with all available bubble chamber's data for CC and NC (anti) neutrino channels over a wide range of neutrino energy.
\begin{acknowledgments}
I would like to thank J. Sobczyk, E. Rondio, P. Przewlocki, C. Wret, K. McFarland, J. Nieves, H. Hayato, J. Zmuda,
R. Gonzalez Jimenez, D. Cherdack, F. Shanchez and C. Wilkinson for the helpful discussions and comments. \\
This work was partially supported by the Polish National Science Centre, project number 2014/14/M/ST2/00850 and Horizon 2020 MSCA-RISE project JENNIFER.
\end{acknowledgments}

\appendix
 \section{Invariant and Isobaric Frame Amplitudes}\label{appA}
 Reference \cite{Adler} provides the following linearly independent Lorentz invariants for vector and axial currents:
\setcounter{equation}{2}
\begin{equation}
 \begin{aligned}
O^{\lambda_k}(V_1)&=\frac{1}{2} \gamma_5\left[ (\gamma e^{\lambda_k})(\gamma k) - (\gamma k)(\gamma e^{\lambda_k})\right]\\
O^{\lambda_k}(V_2)&= -2\gamma_5 \left[ (P e^{\lambda_k})(q k) - (P k)(qe^{\lambda_k})\right]\\
O^{\lambda_k}(V_3)&= \gamma_5 \left[ (\gamma e^{\lambda_k})(q k) - (\gamma k)(qe^{\lambda_k})\right]\\
O^{\lambda_k}(V_4)&= 2\gamma_5 \left[ (\gamma e^{\lambda_k})(P k) - (\gamma k)(Pe^{\lambda_k})\right]\\
&~- M\gamma_5\left[ (\gamma e^{\lambda_k})(\gamma k) - (\gamma k)(\gamma e^{\lambda_k})\right]\\
O^{\lambda_k}(V_5)&= -\gamma_5 \left[ (k e^{\lambda_k})(q k) - k^2(qe^{\lambda_k})\right]\\
O^{\lambda_k}(V_6)&= \gamma_5\left[ (k e^{\lambda_k})(\gamma k) - k^2(\gamma e^{\lambda_k})\right]\\
\\
O^{\lambda_k}(A_1)&= \frac{1}{2}[(\gamma q)(\gamma e^{\lambda_k}) - (\gamma e^{\lambda_k})(\gamma q)] \\ O^{\lambda_k}(A_2)&= 2(e^{\lambda_k} P)\\O^{\lambda_k}(A_3)&= (e^{\lambda_k} q)\\O^{\lambda_k}(A_4)&= M\gamma e^{\lambda_k}\\O^{\lambda_k}(A_5)&= -2(\gamma k)(e^{\lambda_k} P)\\O^{\lambda_k}(A_6)&= - (\gamma k)(e^{\lambda_k} q)\\O^{\lambda_k}(A_7)&=(e^{\lambda_k} k)\\O^{\lambda_k}(A_8)&= -(\gamma k)(e^{\lambda_k} k)
\end{aligned}
\end{equation}
 where $P= 1/2(p_1 + p_2)$ and $e k = \epsilon^ {0} k_0 - \pmb{\epsilon}.\mathbf{k}$.\\
 In the isobaric frame, the following bases are used from Ref.~\cite{Adler}:
\begin{align}
\Sigma_1 &= \pmb{\sigma} \pmb{\epsilon} - (\pmb{\sigma} \hat{\mathbf{k}})(\hat{\mathbf{k}} \pmb{\epsilon})\nonumber\\
\Sigma_2 &= -i(\pmb{\sigma}\hat{\mathbf{q}}) \pmb{\sigma}. (\hat{\mathbf{k}} \times \pmb{\epsilon})\nonumber\\
\Sigma_3 &= (\pmb{\sigma} \hat{\mathbf{k}}) (\hat{\mathbf{q}} \pmb{\epsilon} - (\hat{\mathbf{q}}\hat{\mathbf{k}} )( \hat{\mathbf{k}}\pmb{\epsilon}))\nonumber\\
\Sigma_4 &= (\pmb{\sigma} \hat{\mathbf{q}})  (\hat{\mathbf{q}} \pmb{\epsilon} - (\hat{\mathbf{q}}\hat{\mathbf{k}})( \hat{\mathbf{k}}\pmb{\epsilon}))\nonumber\\
\Sigma_5 &= (\pmb{\sigma} \hat{\mathbf{k}})  \hat{\mathbf{k}}.(k_0\pmb{\epsilon} - \epsilon_0 \mathbf{k})\nonumber\\
\Sigma_6 &= (\pmb{\sigma} \hat{\mathbf{q}})  \hat{\mathbf{k}}.(k_0\pmb{\epsilon} - \epsilon_0\mathbf{k})\nonumber
\end{align}
\begin{align}
 \Lambda_1 &= -\pmb{\sigma} \hat{\mathbf{q}}(\pmb{\sigma} \pmb{\epsilon} - (\pmb{\sigma} \hat{\mathbf{k}})(\hat{\mathbf{k}} \pmb{\epsilon}))\nonumber\\
 \Lambda_2 &= i \pmb{\sigma}.(\hat{\mathbf{k}} \times \pmb{\epsilon})\nonumber\\
 \Lambda_3 &= -(\pmb{\sigma}\hat{\mathbf{q}})(\pmb{\sigma} \hat{k}) (\hat{\mathbf{q}} \pmb{\epsilon} - (\hat{\mathbf{q}}\hat{\mathbf{k}})(\hat{\mathbf{k}}\pmb{\epsilon}))\nonumber\\
 \Lambda_4 &=  -(\hat{\mathbf{q}} \pmb{\epsilon} - (\hat{\mathbf{q}}\hat{\mathbf{k}})  (\hat{\mathbf{k}}\pmb{\epsilon}))\nonumber\\
 \Lambda_5 &= -(\pmb{\sigma}\hat{\mathbf{q}})(\pmb{\sigma} \hat{\mathbf{k}})(\hat{\mathbf{k}} \pmb{\epsilon}) ~\frac{|\mathbf{k}|}{k_0}\nonumber\\
 \Lambda_6 &= -  (\hat{\mathbf{k}}\pmb{\epsilon}) \frac{|\mathbf{k}|}{k_0}\nonumber\\
 \Lambda_7 &= - (\pmb{\sigma}\hat{\mathbf{q}})(\pmb{\sigma} \hat{\mathbf{k}}) (\epsilon k)/k_0\nonumber\\
 \Lambda_8 &= - (\epsilon k)/k_0
 \end{align}
 where
 \begin{equation}\label{}
 \mathbf{A}\pmb{\sigma}= A^i\sigma_{i} = A^1\sigma_1 + A^2\sigma_2+ A^3\sigma_3.
 \end{equation}
 The relation between Lorentz-invariant SPP amplitudes $V_k$, $A_k$ in Eq.~(\ref{Dirac_current}) and $\mathscr{F}_i$, $\mathscr{G}_i$ in Eq.~(\ref{isobar_dicom}) is given in the following way:
\begin{eqnarray}
\mathscr{F}_i &=& K^V_i . F_i ~(i=1, ...,6)\nonumber\\
\mathscr{G}_i &=& K^A_i . G_i ~(i=1, ...,8)\label{FG}
\end{eqnarray}
with
\begin{equation}
\begin{tabular}{ll}
$\begin{aligned}
K_1^V &= W_- O_{1+}\\
K_2^V &= W_+ O_{1-}\\
K_3^V &= q^2 W_+ O_{2-}
\end{aligned}$
&~~
$\begin{aligned}
K_4^V &= q^2 W_+ O_{2-}\\
K_5^V &= 1/O_{2+}\\
K_6^V &= 1/O_{2-}
\end{aligned}$
\end{tabular}
\end{equation}
and
\begin{equation}
\begin{tabular}{ll}
$\begin{aligned}
K_1^A &= |\mathbf{q}| O_{2+}\\
K_2^A &= |\mathbf{q}| O_{2-}\\
K_3^A &= |\mathbf{q}| O_{1-}\\
K_4^A &= |\mathbf{q}| O_{1+}
\end{aligned}$
&~~
$\begin{aligned}
K_5^A &=  O_{1-}\\
K_6^A &= O_{1+}\\
K_7^A &= O_{1-}\\
K_8^A &= O_{1+}
\end{aligned}$
\end{tabular}
\end{equation}
where
\begin{eqnarray}
W_{\pm} &=& W \pm M\nonumber\\
O_{1\pm} &=& \left [(W^2_{\pm} - k^2)(W^2_{\pm} - m_{\pi}^2)\right]^{\frac{1}{2}}/2W\nonumber\\
O_{2\pm} &=& \left [(W^2_{\pm} - k^2)/(W^2_{\pm} - m_{\pi}^2)\right]^{\frac{1}{2}}.
\end{eqnarray}
$F_i$'s for the vector part are
\begin{widetext}
\begin{eqnarray}
\begin{aligned}
F_1&= V_1 + (V_3-V_4)(qk)/W_- +  V_4W_{-} - V_6 k^2/W_-~,\\
F_2&=-V_1 + (V_3-V_4)(qk)/W_+ +  V_4W_{+} - V_6 k^2/W_+~,\\
F_3 &= V_3 - V_4 + V_{25}/W_+ ~, \\
F_4 &= V_3 - V_4 - V_{25}/W_- ~, \\
F_5 &= V_1(W_+^2 - k^2)/2W - V_2(qk)(W_+^2 - k^2 + 2WW_-)/2W + (V_3-V_4)(W_+q_0 - (qk)) \\&+ V_4(W_+^2 - k^2)W_-/2W - V_5(qk)k_0 - V_6 (W_{+}^2 - k^2)W_{-}/2W + q_0 V_{25}~,\\
F_6 &=-V_1(W_-^2 - k^2)/2W + V_2(qk)(W_+^2 - k^2 + 2WW_-)/2W + (V_3-V_4)(W_-q_0 - (qk)) \\&+ V_4(W_-^2 - k^2)W_+/2W + V_5(qk)k_0 - V_6 (W_{-}^2 - k^2)W_{+}/2W - q_0 V_{25}~,
\end{aligned}
\end{eqnarray}
 and $G_i$'s for the axial part of (\ref{FG}) are
\begin{eqnarray}
\begin{aligned}
G_1&= W_+ A_1 - MA_4 ~,\\
G_2&= -W_- A_1- MA_4 ~,\\
G_3 &= A_1 + A_2 - A_3 + (A_5 - A_6)W_{+} ~,\\
G_4 &=-A_1 - A_2 + A_3 + (A_5 - A_6)W_{-} ~,\\
G_5 &= \left[\Delta+ (W_+^2 - m_{\pi}^2)/2W + 2Wk_0 W_+/(W_-^2 - k^2) \right]A_1 + (\Delta + p_{02} + W)A_2 + (q_0 - \Delta)A_3 \\&- M\left[ W_{-} /(p_{01}- M)\right]A_4 + W_{+} \left[ (\Delta + p_{02} + W)A_5 + (q_0 - \Delta)A_6\right] ~,  \\
G_6 &= -\left[\Delta+ (W_-^2 - m_{\pi}^2)/2W + 2Wk_0 W_-/(W_+^2 - k^2) \right]A_1 + (\Delta + p_{02} + W)A_2 - (q_0 - \Delta)A_3 \\&- M\left[ W_{+} /(p_{01}+ M)\right]A_4 + W_{-} \left[ (\Delta + p_{02} + W)A_5 + (q_0 - \Delta)A_6\right] ~,  \\
G_7 &= (W_+^2 - m_{\pi}^2)A_1/2W + (p_{01} + p_{02})A_2 + q_0 A_3 - MA_4 + k_0A_7 \\&+ W_+\left[ (p_{01} + p_{02})A_5 + q_0A_6 + k_0A_8\right] ~,\\
G_8 &=-(W_-^2 - m_{\pi}^2)A_1/2W - (p_{01} + p_{02})A_2 - q_0 A_3 - MA_4 - k_0A_7 \\&+ W_-\left[  (p_{01} + p_{02})A_5 + q_0A_6 + k_0A_8\right] ~,
\end{aligned}
\end{eqnarray}
\end{widetext}
where
\begin{eqnarray}
V_{25}&=& W_+ W_- V_2 + k^2 V_5\nonumber\\
\Delta &=& k_0(q_0k_0 -(qk))/ \bf{k}^2.
\end{eqnarray}
\section{Helicity Amplitudes}\label{HAapp}
According to the isobaric frame which is shown in Fig.~\ref{Isoframe}, the momentum vectors are:
\begin{eqnarray}
\mathbf{k}&=& |\mathbf{k}|\begin{pmatrix} 0&0&1 \end{pmatrix}\nonumber\\
\mathbf{q}&=& |\mathbf{q}|\begin{pmatrix} \sin\theta \cos\phi&\sin\theta \sin\phi&cos\theta \end{pmatrix}~,\nonumber\\
\end{eqnarray}
and the nucleon spinors are:
\begin{eqnarray}
\chi_1(\uparrow) = \begin{pmatrix} 0\\1 \end{pmatrix},~~~~~  \chi_1(\downarrow) = \begin{pmatrix} -1\\0 \end{pmatrix}
\end{eqnarray}
\begin{eqnarray}
\chi_2(\uparrow) = \begin{pmatrix} \sin \theta/2\\-e^{i\phi} \cos\theta/2 \end{pmatrix},~  \chi_2(\downarrow) = \begin{pmatrix} e^{-i\phi} \cos\theta/2\\ \sin \theta/2 \end{pmatrix}~.\nonumber\\
\end{eqnarray}
Using Eq.~(\ref{HA_definition}), the helicity amplitudes in the isobaric frame can be obtained which are displayed in Table~\ref{helicity_amp}.\\
\begin{turnpage}
\begin{table*}
\centering
\caption{Helicity amplitudes}
\label{helicity_amp}
\renewcommand{\arraystretch}{1.3}
\begin{ruledtabular}
 \begin{tabular}{ll}
 Hadronic vector current  & Hadronic Axial vector current   \\ [0.5ex]
 \hline
$\begin{aligned}
&\scalebox{0.001}{~}\\
\tilde{F}_{\frac{1}{2} \frac{1}{2}}^{e_L} &= \frac{1}{\sqrt{2}} e^{-2i\phi} \sin{\theta} \cos\frac{\theta}{2} (\mathscr{F}_3 + \mathscr{F}_4)\nonumber\\
\tilde{F}_{\frac{-1}{2} \frac{1}{2}}^{e_L} &= -\frac{1}{\sqrt{2}} e^{-i\phi} \sin{\theta} \sin\frac{\theta}{2} (\mathscr{F}_3 - \mathscr{F}_4)\nonumber\\
\tilde{F}_{\frac{1}{2} \frac{-1}{2}}^{e_L} &= \sqrt{2} e^{-i\phi} \big[\cos\frac{\theta}{2} (\mathscr{F}_1 - \mathscr{F}_2) - \frac{1}{2} \sin{\theta} \sin\frac{\theta}{2} (\mathscr{F}_3 - \mathscr{F}_4)\big]\nonumber\\
\tilde{F}_{\frac{-1}{2} \frac{-1}{2}}^{e_L} &= -\sqrt{2}  \big[\sin\frac{\theta}{2} (\mathscr{F}_1 + \mathscr{F}_2) + \frac{1}{2} \sin{\theta} \cos\frac{\theta}{2} (\mathscr{F}_3 + \mathscr{F}_4)\big]\\[3pt]
\end{aligned}$
&
$\begin{aligned}
&\scalebox{0.001}{~}\\
 \tilde{G}_{\frac{1}{2} \frac{1}{2}}^{e_L} &= \frac{1}{\sqrt{2}} e^{-2i\phi} \sin{\theta} \cos\frac{\theta}{2} (\mathscr{G}_3 + \mathscr{G}_4)\nonumber\\
\tilde{G}_{\frac{-1}{2} \frac{1}{2}}^{e_L} &= \frac{1}{\sqrt{2}} e^{-i\phi} \sin{\theta} \sin\frac{\theta}{2} (\mathscr{G}_3 - \mathscr{G}_4)\nonumber\\
\tilde{G}_{\frac{1}{2} \frac{-1}{2}}^{e_L} &= \sqrt{2} e^{i\phi} \big[\cos\frac{\theta}{2} (\mathscr{G}_1 - \mathscr{G}_2) - \frac{1}{2} \sin{\theta} \sin\frac{\theta}{2} (\mathscr{G}_3 - \mathscr{G}_4)\big]\nonumber\\
\tilde{G}_{\frac{-1}{2} \frac{-1}{2}}^{e_L} &= \sqrt{2}  \big[\sin\frac{\theta}{2} (\mathscr{G}_1 + \mathscr{G}_2) + \frac{1}{2} \sin{\theta} \cos\frac{\theta}{2} (\mathscr{G}_3 + \mathscr{G}_4)\big]\\[3pt]
 \end{aligned}$
 \\ \hline
$\begin{aligned}
&\scalebox{0.001}{~}\\
\tilde{F}_{\frac{1}{2} \frac{1}{2}}^{e_R} &= -\sqrt{2}  \big[\sin\frac{\theta}{2} (\mathscr{F}_1 + \mathscr{F}_2) + \frac{1}{2} \sin{\theta} \cos\frac{\theta}{2} (\mathscr{F}_3 + \mathscr{F}_4)\big]\nonumber\\
\tilde{F}_{\frac{-1}{2} \frac{1}{2}}^{e_R} &= -\sqrt{2} e^{i\phi} \big[\cos\frac{\theta}{2} (\mathscr{F}_1 - \mathscr{F}_2) - \frac{1}{2} \sin{\theta} \sin\frac{\theta}{2} (\mathscr{F}_3 - \mathscr{F}_4)\big]\nonumber\\
\tilde{F}_{\frac{1}{2} -\frac{1}{2}}^{e_R} &= \frac{1}{\sqrt{2}} e^{i\phi} \sin{\theta} \sin\frac{\theta}{2} (\mathscr{F}_3 - \mathscr{F}_4)\nonumber\\
\tilde{F}_{-\frac{1}{2} -\frac{1}{2}}^{e_R} &= \frac{1}{\sqrt{2}} e^{2i\phi} \sin{\theta} \cos\frac{\theta}{2} (\mathscr{F}_3 + \mathscr{F}_4)\\[3pt]
\end{aligned}$
&
$\begin{aligned}
&\scalebox{0.001}{~}\\
 \tilde{G}_{\frac{1}{2} \frac{1}{2}}^{e_R} &= -\sqrt{2}  \big[\sin\frac{\theta}{2} (\mathscr{G}_1 + \mathscr{G}_2) + \frac{1}{2} \sin{\theta} \cos\frac{\theta}{2} (\mathscr{G}_3 + \mathscr{G}_4)\big]\nonumber\\
\tilde{G}_{-\frac{1}{2} \frac{1}{2}}^{e_R} &= \sqrt{2} e^{i\phi} \big[\cos\frac{\theta}{2} (\mathscr{G}_1 - \mathscr{G}_2) - \frac{1}{2} \sin{\theta} \sin\frac{\theta}{2} (\mathscr{G}_3 - \mathscr{G}_4)\big]\nonumber\\
\tilde{G}_{\frac{1}{2} -\frac{1}{2}}^{e_R} &= \frac{1}{\sqrt{2}} e^{i\phi} \sin{\theta} \sin\frac{\theta}{2} (\mathscr{G}_3 - \mathscr{G}_4)\nonumber\\
\tilde{G}_{-\frac{1}{2} -\frac{1}{2}}^{e_R} &=  -\frac{1}{\sqrt{2}} e^{2i\phi} \sin{\theta} \cos\frac{\theta}{2} (\mathscr{G}_3 + \mathscr{G}_4)\\[3pt]
 \end{aligned}$
 \\ \hline
$\begin{aligned}
&\scalebox{0.001}{~}\\
\tilde{F}_{\frac{1}{2} \frac{1}{2}}^{e_{-}} &= e^{-i\phi}\cos\frac{\theta}{2}\frac{1}{C_{-}}(k_0\epsilon^0_L - |\mathbf{k}| \epsilon^3_L )(\mathscr{F}_5 + \mathscr{F}_6)\nonumber \\
\tilde{F}_{-\frac{1}{2} \frac{1}{2}}^{e_{-}} &=  -\sin\frac{\theta}{2}\frac{1}{C_-}(k_0\epsilon^0_L - |\mathbf{k}| \epsilon^3_L )(\mathscr{F}_5 - \mathscr{F}_6) \nonumber\\
\tilde{F}_{\frac{1}{2} -\frac{1}{2}}^{e_{-}} &= -\sin\frac{\theta}{2}\frac{1}{C_-}(k_0\epsilon^0_L - |\mathbf{k}| \epsilon^3_L)(\mathscr{F}_5 - \mathscr{F}_6) \nonumber\\
\tilde{F}_{-\frac{1}{2} -\frac{1}{2}}^{e_{-}} &= - e^{i\phi}\cos\frac{\theta}{2}\frac{1}{C_-}(k_0\epsilon^0_L - |\mathbf{k}| \epsilon^3_L)(\mathscr{F}_5 + \mathscr{F}_6)\\[3pt]
\end{aligned}$
&
$\begin{aligned}
&\scalebox{0.001}{~}\\
\tilde{G}_{\frac{1}{2} \frac{1}{2}}^{e_{-}} &= e^{-i\phi}\cos\frac{\theta}{2}\frac{1}{C_- k_0}  \big[|\mathbf{k}| \epsilon^3_L(\mathscr{G}_5 + \mathscr{G}_6) +  (k_0\epsilon^0_L - |\mathbf{k}| \epsilon^3_L)(\mathscr{G}_7 + \mathscr{G}_8)\big]\nonumber\\
\tilde{G}_{-\frac{1}{2} \frac{1}{2}}^{e_{-}} &= \sin\frac{\theta}{2}\frac{1}{C_- k_0}  \big[|\mathbf{k}| \epsilon^3_L (\mathscr{G}_5 - \mathscr{G}_6) +  (k_0\epsilon^0_L - |\mathbf{k}| \epsilon^3_L)(\mathscr{G}_7 - \mathscr{G}_8)\big]\nonumber\\
\tilde{G}_{\frac{1}{2} -\frac{1}{2}}^{e_{-}} &= -\sin\frac{\theta}{2}\frac{1}{C_- k_0}  \big[|\mathbf{k}| \epsilon^3_L(\mathscr{G}_5 - \mathscr{G}_6) +  (k_0\epsilon^0_L - |\mathbf{k}| \epsilon^3_L)(\mathscr{G}_7 - \mathscr{G}_8)\big]\nonumber\\
\tilde{G}_{-\frac{1}{2} -\frac{1}{2}}^{e_{-}} &= e^{i\phi}\cos\frac{\theta}{2}\frac{1}{C_- k_0}  \big[|\mathbf{k}| \epsilon^3_L(\mathscr{G}_5 + \mathscr{G}_6) +  (k_0\epsilon^0_L - |\mathbf{k}| \epsilon^3_L)(\mathscr{G}_7 + \mathscr{G}_8)\big]\\[3pt]
 \end{aligned}$
 \\ \hline
$\begin{aligned}
&\scalebox{0.001}{~}\\
\tilde{F}_{\frac{1}{2} \frac{1}{2}}^{e_{+}} &= e^{-i\phi}\cos\frac{\theta}{2}\frac{1}{C_{+}}(k_0\epsilon^0_R - |\mathbf{k}| \epsilon^3_R)(\mathscr{F}_5 + \mathscr{F}_6) \nonumber\\
\tilde{F}_{-\frac{1}{2} \frac{1}{2}}^{e_{+}} &=  -\sin\frac{\theta}{2}\frac{1}{C_+}(k_0\epsilon^0_R - |\mathbf{k}| \epsilon^3_R)(\mathscr{F}_5 - \mathscr{F}_6) \nonumber\\
\tilde{F}_{\frac{1}{2} -\frac{1}{2}}^{e_{+}} &= -\sin\frac{\theta}{2}\frac{1}{C_+}(k_0\epsilon^0_R - |\mathbf{k}| \epsilon^3_R)(\mathscr{F}_5 - \mathscr{F}_6) \nonumber\\
\tilde{F}_{-\frac{1}{2} -\frac{1}{2}}^{e_{+}} &= - e^{i\phi}\cos\frac{\theta}{2}\frac{1}{C_+}(k_0\epsilon^0_R - |\mathbf{k}| \epsilon^3_R)(\mathscr{F}_5 + \mathscr{F}_6)\\[3pt]
\end{aligned}$
&
$\begin{aligned}
&\scalebox{0.001}{~}\\
 \tilde{G}_{\frac{1}{2} \frac{1}{2}}^{e_{+}} &= e^{-i\phi}\cos\frac{\theta}{2}\frac{1}{C_+ k_0}  \big[|\mathbf{k}| \epsilon^3_R(\mathscr{G}_5 + \mathscr{G}_6) +  (k_0\epsilon^0_R - |\mathbf{k}| \epsilon^3_R)(\mathscr{G}_7 + \mathscr{G}_8)\big]\nonumber\\
\tilde{G}_{-\frac{1}{2} \frac{1}{2}}^{e_{+}} &= \sin\frac{\theta}{2}\frac{1}{C_+ k_0}  \big[|\mathbf{k}| \epsilon^3_R(\mathscr{G}_5 - \mathscr{G}_6) +  (k_0\epsilon^0_R - |\mathbf{k}| \epsilon^3_R)(\mathscr{G}_7 - \mathscr{G}_8)\big]\nonumber\\
\tilde{G}_{\frac{1}{2} -\frac{1}{2}}^{e_{+}} &= -\sin\frac{\theta}{2}\frac{1}{C_+ k_0}  \big[|\mathbf{k}| \epsilon^3_R(\mathscr{G}_5 - \mathscr{G}_6) +  (k_0\epsilon^0_R - |\mathbf{k}| \epsilon^3_R)(\mathscr{G}_7 - \mathscr{G}_8)\big]\nonumber\\
\tilde{G}_{-\frac{1}{2} -\frac{1}{2}}^{e_{+}} &= e^{i\phi}\cos\frac{\theta}{2}\frac{1}{C_+ k_0}  \big[|\mathbf{k}| \epsilon^3_R(\mathscr{G}_5 + \mathscr{G}_6) +  (k_0\epsilon^0_R - |\mathbf{k}| \epsilon^3_R)(\mathscr{G}_7 + \mathscr{G}_8)\big]\\[3pt]
 \end{aligned}$
 \\
\end{tabular}
\end{ruledtabular}
\end{table*}
\end{turnpage}
The explicit form of the $d^j_{\lambda,\mu}(\theta)$ functions for $j=l+\frac{1}{2}$ are given in Eq.~(\ref{dj_def}):
\begin{align}
d^j_{\frac{1}{2} \frac{1}{2}}~&= (l+1)^{-1} \cos\frac{\theta}{2} (P'_{l+1} - P'_l)\nonumber\\
d^j_{-\frac{1}{2} \frac{1}{2}}&= (l+1)^{-1} \sin\frac{\theta}{2} (P'_{l+1} + P'_l)\nonumber\\
d^j_{\frac{1}{2} \frac{3}{2}}~&= (l+1)^{-1} \sin\frac{\theta}{2} (\sqrt{\frac{l}{l+2}}P'_{l+1} + \sqrt{\frac{l+2}{l}} P'_l)\nonumber\\
d^j_{-\frac{1}{2} \frac{3}{2}}&= (l+1)^{-1} \cos\frac{\theta}{2} (-\sqrt{\frac{l}{l+2}}P'_{l+1} + \sqrt{\frac{l+2}{l}} P'_l)
\label{dj_def}
\end{align}
where $P_l$ are Legendre polynomials and $P'_l= dP_l/d\cos\theta$.\\
\section{Differential cross section}\label{xsec_app}
Equation (\ref{Xsec}) can be expanded by using Table \ref{helicity_amp}:
\begin{widetext}
\begin{eqnarray} \label{Xsec_phi}
\frac{d\sigma(\nu N  \rightarrow lN\pi)}{dk^2 dW d\Omega_{\pi}}  &=&
\frac{G_F^2}{2} \frac{1}{(2\pi)^4} \frac{|\bf{q}|}{4} \frac{-k^2}{(k^L)^2}\sum_{\lambda_2, \lambda_1} \Bigg\{\nonumber\\
&~& |C_{L}|^2 |\tilde{F}_{\lambda_2 \lambda_1}^{e_L}(\theta)- \tilde{G}_{\lambda_2 \lambda_1}^{e_L}(\theta)|^2
+ |C_{R}|^2 |\tilde{F}_{\lambda_2 \lambda_1}^{e_R}(\theta)- \tilde{G}_{\lambda_2 \lambda_1}^{e_R}(\theta)|^2\nonumber\\
&+& |C_{-}|^2 ~|\tilde{F}_{\lambda_2 \lambda_1}^{e_{-}}(\theta) - \tilde{G}_{\lambda_2 \lambda_1}^{e_{-}}(\theta)|^2
+ |C_{+}|^2 ~|\tilde{F}_{\lambda_2 \lambda_1}^{e_+}(\theta)- \tilde{G}_{\lambda_2 \lambda_1}^{e_+}(\theta)|^{2}\nonumber\\
+ 2\cos\phi \Big\{
&~& C_{L_{-}} C_{-} ~\Re \left[ (\tilde{F}_{\lambda_2 \lambda_1}^{e_L}(\theta)- \tilde{G}_{\lambda_2 \lambda_1}^{e_L}(\theta))^{\ast}
                                       (\tilde{F}_{\lambda_2 \lambda_1}^{e_{-}}(\theta)- \tilde{G}_{\lambda_2 \lambda_1}^{e_{-}}(\theta))\right] \nonumber\\
&+& C_{R_{-}} C_{-} ~\Re \left[ (\tilde{F}_{\lambda_2 \lambda_1}^{e_R}(\theta)- \tilde{G}_{\lambda_2 \lambda_1}^{e_R}(\theta))^{\ast}
                                       (\tilde{F}_{\lambda_2 \lambda_1}^{e_{-}}(\theta)- \tilde{G}_{\lambda_2 \lambda_1}^{e_{-}}(\theta))\right]  \nonumber\\
&+& C_{L_{+}} C_{+} ~\Re \left[ (\tilde{F}_{\lambda_2 \lambda_1}^{e_L}(\theta)- \tilde{G}_{\lambda_2 \lambda_1}^{e_L}(\theta))^{\ast}
                                       (\tilde{F}_{\lambda_2 \lambda_1}^{e_{+}}(\theta)- \tilde{G}_{\lambda_2 \lambda_1}^{e_{+}}(\theta))\right]  \nonumber\\
&+& C_{R_{+}} C_{+} ~\Re \left[ (\tilde{F}_{\lambda_2 \lambda_1}^{e_R}(\theta)- \tilde{G}_{\lambda_2 \lambda_1}^{e_R}(\theta))^{\ast}
                                       (\tilde{F}_{\lambda_2 \lambda_1}^{e_{+}}(\theta)- \tilde{G}_{\lambda_2 \lambda_1}^{e_{+}}(\theta))\right]   \Big\} \nonumber\\
+ 2\sin\phi \Big\{
&-&  C_{L_{-}} C_{-} ~\Im \left[ (\tilde{F}_{\lambda_2 \lambda_1}^{e_L}(\theta)- \tilde{G}_{\lambda_2 \lambda_1}^{e_L}(\theta))^{\ast}
                                       (\tilde{F}_{\lambda_2 \lambda_1}^{e_{-}}(\theta)- \tilde{G}_{\lambda_2 \lambda_1}^{e_{-}}(\theta))\right] \nonumber\\
&+& C_{R_{-}} C_{-} ~\Im \left[ (\tilde{F}_{\lambda_2 \lambda_1}^{e_R}(\theta)- \tilde{G}_{\lambda_2 \lambda_1}^{e_R}(\theta))^{\ast}
                                       (\tilde{F}_{\lambda_2 \lambda_1}^{e_{-}}(\theta)- \tilde{G}_{\lambda_2 \lambda_1}^{e_{-}}(\theta))\right]  \nonumber\\
&-& C_{L_{+}} C_{+} ~\Im \left[ (\tilde{F}_{\lambda_2 \lambda_1}^{e_L}(\theta)- \tilde{G}_{\lambda_2 \lambda_1}^{e_L}(\theta))^{\ast}
                                (\tilde{F}_{\lambda_2 \lambda_1}^{e_{+}}(\theta)- \tilde{G}_{\lambda_2 \lambda_1}^{e_{+}}(\theta))\right]  \nonumber\\
&+& C_{R_{+}} C_{+} ~\Im \left[ (\tilde{F}_{\lambda_2 \lambda_1}^{e_R}(\theta)- \tilde{G}_{\lambda_2 \lambda_1}^{e_R}(\theta))^{\ast}
                                       (\tilde{F}_{\lambda_2 \lambda_1}^{e_{+}}(\theta)- \tilde{G}_{\lambda_2 \lambda_1}^{e_{+}}(\theta))\right]   \Big\} \nonumber\\
+ 2\cos2\phi (C_{L_{-}}^{\ast}C_{R_{-}} &+& C_{L_{+}}^{\ast}C_{R_{+}}) \Re\left[ (\tilde{F}_{\lambda_2 \lambda_1}^{e_L}(\theta)- \tilde{G}_{\lambda_2 \lambda_1}^{e_L}(\theta))^{\ast}(\tilde{F}_{\lambda_2 \lambda_1}^{e_R}(\theta)- \tilde{G}_{\lambda_2 \lambda_1}^{e_R}(\theta))\right] \nonumber\\
 -2\sin2\phi (C_{L_{-}}^{\ast}C_{R_{-}} &+& C_{L_{+}}^{\ast}C_{R_{+}}) \Im \left[ (\tilde{F}_{\lambda_2 \lambda_1}^{e_L}(\theta)- \tilde{G}_{\lambda_2 \lambda_1}^{e_L}(\theta))^{\ast}(\tilde{F}_{\lambda_2 \lambda_1}^{e_R}(\theta)- \tilde{G}_{\lambda_2 \lambda_1}^{e_R}(\theta))\right] \Bigg\}\nonumber\\
\end{eqnarray}
\end{widetext}
where
\begin{equation}\label{CTpm}
|C_{L(R)}|^2 = |C_{L(R)_-}|^2 + |C_{L(R)_+}|^2.
\end{equation}

\nocite{*}
\bibliography{apssamp}

\end{document}